\documentclass[11pt]{article}
\usepackage{hyperref}
\usepackage{algorithm,algorithmic}
\usepackage{amsfonts,amsmath, amssymb, bm}
\usepackage{natbib}% for citations
\def\math#1{$#1$}

\def\frac#1#2{{#1\over #2}}

\def\prob#1{{P\left[{#1}\right]}}

\def\x{{\mathbf x}}
\def\y{{\mathbf y}}

\def\dotfil{\leaders\hbox to 1.5mm{.}\hfill}
\newcounter{rmnum}
\def\RN#1{\setcounter{rmnum}{#1}\uppercase\expandafter{\romannumeral\value{rmnum}}}
\def\rn#1{\setcounter{rmnum}{#1}\expandafter{\romannumeral\value{rmnum}}}

\newcommand{\ONorm }[1]{\mbox{}\left\|#1\right\|_1  }
\newcommand{\ONormS}[1]{\mbox{}\left\|#1\right\|_1^2}
\newcommand{\ZNorm }[1]{\mbox{}\left\|#1\right\|_0  }

\newcommand{\FNorm }[1]{\mbox{}\left\|#1\right\|_F  }
\newcommand{\FNormS}[1]{\mbox{}\left\|#1\right\|_F^2}

\newcommand{\TNorm}[1]{\mbox{}\left\|#1\right\|_2}
\newcommand{\TNormS}[1]{\mbox{}\left\|#1\right\|_2^2}
\newcommand{\TsNorm}[1]{\mbox{}\|#1\|_2}

\newcommand{\XsNorm }[1]{\mbox{}\|#1\|_{\xi}  }

\newcommand{\setlinespacing}[1]%
           {\setlength{\baselineskip}{#1 \defbaselineskip}}

\newcommand{\rank}[1]{{\bf rank}{\left(#1\right)}}
\newcommand{\abs }[1]{\left|#1\right|}

\newtheorem{lemma}{Lemma}
\newtheorem{theorem}{Theorem}

\newenvironment{Proof}{\noindent {\em Proof:}}{\\\hspace*{\fill}\mbox{$\diamond$}}

\newcommand{\mat}[1]{{\ensuremath{\bm{\mathrm{#1}}}}}

\def\exp{\hbox{\rm exp}}
\def\rank{\hbox{\rm rank}}

\def\e{{\mathbf e}}

\def\x{{\mathbf x}}

\def\matA{\mat{A}}

\def\matD{\mat{D}}

\def\matG{\mat{G}}

\def\matI{\mat{I}}

\def\matM{\mat{M}}

\def\matR{\mat{R}}

\def\matU{\mat{U}}
\def\matV{\mat{V}}

\def\matX{\mat{X}}
\def\matY{\mat{Y}}

\def\matSig{\mat{\Sigma}}

\usepackage{graphicx}
\usepackage{epstopdf}
\usepackage{float}
\usepackage{caption}
\usepackage{subcaption}
\usepackage{times}
\usepackage{multirow}
\begin{document}
%\vskip 0.3in
\title{Recovering PCA from Hybrid-$(\ell_1,\ell_2)$ Sparse Sampling of
Data Elements}
\author{
Abhisek Kundu
\thanks{
Department of Computer Science,
Rensselaer Polytechnic Institute,
Troy, NY,
kundua2@rpi.edu.
}
\and
Petros Drineas
\thanks{
Department of Computer Science,
Rensselaer Polytechnic Institute,
Troy, NY,
drinep@cs.rpi.edu.
}
\and
Malik Magdon-Ismail
\thanks{
Department of Computer Science,
Rensselaer Polytechnic Institute,
Troy, NY,
magdon@cs.rpi.edu.
}
}
\date{}
\maketitle
\begin{abstract}
\noindent 
This paper addresses how well we can recover a data
matrix when only given a few of its elements.
We present a randomized algorithm that 
element-wise sparsifies the data, 
retaining only a few its elements. Our new algorithm
independently samples the data using sampling probabilities 
that depend on both the squares ($\ell_2$ sampling) %\cite{AM01, DZ11}) 
and absolute values ($\ell_1$ sampling) %\cite{AM01, AKL13}) 
of the entries. 
We prove that the hybrid algorithm recovers a near-PCA
reconstruction of the data from a sublinear sample-size:
hybrid-($\ell_1,\ell_2$)  inherits the \math{\ell_2}-ability 
to sample the important elements
as well as the regularization properties of $\ell_1$ sampling, and 
gives strictly better performance than either \math{\ell_1} or \math{\ell_2}
on their own. We also give a one-pass version of our algorithm and 
show experiments to corroborate the theory.
\end{abstract}

\def\norm#1{\|#1\|}
\def\R{{\mathbb{R}}}
\newcommand{\remove}[1]{}
\section{Introduction}

We address the problem of recovering a near-PCA reconstruction of
the data from just a few of its entries --
element-wise matrix sparsification (\cite{AM01, AM07}). Read: you have a 
small sample of data points and those data points have missing features.
This is a situation that one is confronted 
with all too often in machine learning. For example, with 
user-recommendation data, one does not have all the ratings of 
any given user. Or in a privacy preserving setting, a client
may not want to give you all entries in the data matrix.
In such a setting, our goal is to show that if the samples that you 
do get are chosen carefully, the top-\math{k} PCA 
features of the data can be recovered within some provable error bounds.

More formally,
the data matrix is  
$\matA \in \mathbb{R}^{m \times n}$ (\math{m} data points in 
\math{n} dimensions). Often, real data matrices have low effective rank,
so let \math{\matA_k} be the best rank-\math{k} approximation to 
\math{\matA} with \math{\TsNorm{\matA-\matA_k}} being small. \math{\matA_k}
is obtained by projecting \math{\matA} onto the subspace spanned by its
top-\math{k} principal components. In order to approximate this top-\math{k}
principal subspace, we adopt the following strategy.
Select a small number, $s$, of elements from $\matA$ 
and produce a sparse sketch $\tilde{\matA}$; use the sparse sketch
\math{\tilde\matA} to approximate the top-\math{k} singular 
subspace. In Section \ref{sec:PCA}, we give the details of the
 algorithm and the theoretical
guarantees on how well we recover the top-\math{k} principal subspace.
The key quantity that one must control to 
recover a close approximation to PCA is
how well the sparse sketch approximates the data \emph{in the operator norm}.
That is, if \math{\norm{\matA-\tilde\matA}_2} is small then you can
recover PCA effectively.
%--------------------------------------------
\begin{center}
\fbox{
\begin{minipage}{1.0\textwidth}
\underline{{\bf Problem: sparse sampling of data elements}}
\\[0.1in]
Given \math{\matA\in\R^{m\times n}} and \math{\epsilon>0},
sample a small number of elements \math{s} to obtain a 
sparse sketch \math{\tilde\matA} for which
\begin{eqnarray}\label{eqn:sparsification}
\norm{\matA - \tilde{\matA}}_2 \leq \epsilon &\text{ and }& 
\norm{\tilde\matA}_0 \leq s.
\end{eqnarray}
\end{minipage}
}
\end{center}
%--------------------------------------------
Our main result addresses the problem above. In a nutshell,
with only partially observed data that have been 
carefully selected, one can recover an approximation to the 
top-\math{k} principal subspace. An additional benefit is that 
computing our approximation to the top-\math{k} subspace 
using iterated multiplication
can benefit computationally from sparsity.
To construct \math{\tilde \matA}, we use a general 
randomized approach which independently
samples (and rescales) $s$ 
elements from $\matA$ using probability 
$p_{ij}$ to sample element $\matA_{ij}$. We analyze in detail 
the case \math{p_{ij}\propto \alpha|\matA_{ij}| + (1-\alpha)|\matA_{ij}|^2} to get a bound
on \math{\norm{\matA-\tilde\matA}_2}.
We now make our discussion precise, starting with our notation.
\subsection{Notation}\label{sec:notation}
We use bold uppercase (e.g., $\matX$) for
matrices and bold lowercase (e.g., $\x$) for column vectors. 
The $i$-th row of $\matX$ is $\matX_{(i)}$, and the $i$-th column of $\matX$ 
is $\matX^{(i)}$. Let $[n]$ denote the set $\{1, 2, ..., n\}$. 
$\mathbb{E}(X)$ is the expectation of a random variable $X$; 
for a matrix, $\mathbb{E}(\matX)$ denotes the element-wise expectation. For a matrix $\matX \in \mathbb{R}^{m \times n}$, the Frobenius norm $\FNorm{\matX}$ is 
$\FNormS{\matX} = \sum_{i,j=1}^{m,n} \matX_{ij}^2$, and the spectral (operator)
 norm $\TNorm{\matX}$ is 
$\TNorm{\matX} = \text{max}_{\TNorm{\y}=1}\TNorm{\matX\y}$. 
We also have the \math{\ell_1} and \math{\ell_0} norms:
$\ONorm{\matX} = \sum_{i,j=1}^{m,n} \abs{\matX_{ij}}$ and $\ZNorm{\matX}$ 
(the number of non-zero entries in $\matX$). The $k$-th largest singular value of $\matX$ is $\sigma_k(\matX)$.
For symmetric matrices \math{\matX,\ \matY}, \math{\matY \succeq \matX} 
if and only if $\matY - \matX$ is positive semi-definite. $\matI_n$ is the $n \times n$ identity
and $\ln x$ is the natural logarithm of $x$. We use $\e_i$ to denote standard basis vectors whose dimensions will be clear from the context.

Two popular sampling schemes are $\ell_1$ ($p_{ij} = {\abs{\matA_{ij}}}/{\ONorm{\matA}}$ \cite{AM01,AKL13}) and 
$\ell_2$ ($p_{ij} = {\matA_{ij}^2}/{\FNormS{\matA}}$ \cite{AM01, DZ11}). 
We construct $\tilde{\matA}$ as follows: 
$\tilde{\matA}_{ij} = 0$ if the $(i,j)$-th entry is not sampled;
sampled elements $\matA_{ij}$ are rescaled to 
$\tilde{\matA}_{ij} = \matA_{ij}/p_{ij}$ which makes the 
sketch $\tilde{\matA}$ an unbiased estimator of $\matA$, so
$\mathbb{E}[\tilde{\matA}] = \matA$. The sketch is \emph{sparse} if the 
number of sampled elements is sublinear, \math{s=o(mn)}.
Sampling according to element magnitudes is natural in many applications,
for example in a recommendation system users tend to rate a product they 
either like (high positive) or dislike (high negative). 
%This effectively gives us a sparse picture of user-product relations. 
%
%Similarly, in a network matrix (social or biological) where a node represents an entity and an edge between a pair of nodes represents some association between them, most of the edges are non-existent due to lack of interaction.
%Thus, we expect $\tilde{\matA}$ to approximate some properties of $\matA$ (the rescaling step of construction of $\tilde{\matA}$ is mainly mathematical).

Our main sparsification algorithm (Algorithm \ref{alg:alg1}) receives as input a matrix $\matA$ and an accuracy parameter $\epsilon > 0$, and samples $s$ elements from $\matA$ in $s$ independent, identically distributed trials with replacement, according to a hybrid-$(\ell_1, \ell_2)$ probability distribution specified  
in equation
(\ref{eqn:probability}).
The algorithm returns $\tilde{\matA} \in \mathbb{R}^{m \times n}$,  a sparse and unbiased estimator of $\matA$, as a solution to (\ref{eqn:sparsification}). 

\remove{

%
%=================================
\subsection{Motivation}
%=================================
%
\textit{Faster Computation of Low-rank Approximation:}
Element-wise matrix sparsification was pioneered by~\cite{AM01,AM07} for the purpose of speeding up the computation of low-rank approximations of matrices. \cite{AM01,AM07} interpreted matrix sparsification as addition of a zero-mean, random noise matrix $\matR$ to $\matA$. It was then argued that strong linear trends (as captured, for example, by the top few singular vectors) in the ideal data matrix $\matA$, would likely remain prominent in the noisy data $\matA + \matR$, as long as, $\matR$ does not have any meaningful spectral structure to obscure the linear trend of $\matA$. This is because the energy of $\matR$ would likely be distributed uniformly among the top few singular vectors, thus affecting them almost equally. 
Then, computing an appropriate low-rank approximation of $\tilde{\matA} = \matA + \matR$ would be a reasonable surrogate for a low-rank approximation of $\matA$. Moreover, computation of a low-rank approximation of $\tilde{\matA}$ would be faster than that of  $\matA$ because of the sparsity of $\tilde{\matA}$. \cite{AM01, AM07} showed that
\begin{eqnarray}\label{eqn:sparsification_rank_k}
\TNorm{\matA - \tilde{\matA}_k} &\leq& \TNorm{\matA - {\matA}_k} + error, \\
\nonumber error \text{     }&\propto& \text{     }  \TNorm{\matA - \tilde{\matA}}.
\end{eqnarray}
Here $\matA_k$ (respectively $\tilde{\matA}_k$) is the best rank $k$ approximation of $\matA$ (respectively $\tilde{\matA}$) as computed via the Singular Value Decomposition (SVD). Note that, $\matA_k$ is the best rank-$k$ approximation of $\matA$, in spectral norm, i.e., $\TNorm{\matA - {\matA}_k}$ is a solution to the following optimization problem: 
$\min\TNorm{\matA - \matD}$ over all $\matD$, such that $\rank(\matD)\leq k$.
%
%minimum error we can achieve  for any rank-$k$ approximation of $\matA$, in spectral norm.\\ 
%
Thus, (\ref{eqn:sparsification_rank_k}) gives us a connection between the ideal error and the error due to sparse approximation. 
In (\ref{eqn:sparsification_rank_k}), $\tilde{\matA}-\matA$ can be interpreted as the zero-mean noise matrix $\matR$ that is added to $\matA$ to produce sparse data $\tilde{\matA}$.
Naturally, we want to find an unbiased estimator $\tilde{\matA}$ that controls $\TNorm{\matA - \tilde{\matA}}$ (spectral structure of noise matrix) to preserve spectral structure of $\matA$ in $\tilde{\matA}$, while maintaining sparsity of $\tilde{\matA}$ for computational gain. This leads us to the  optimization problem in (\ref{eqn:sparsification}). \\
\\
\textit{Faster Computation of Principal Component Analysis (PCA):}
PCA is a fundamental task in data analysis in order to represent the data succinctly in a low dimensional space spanned by top few principal components. One practical hurdle of PCA is its computation time as it slows down considerably for large datasets. We can use element-wise sparsification as a heuristic to speed up the computation of (approximate) PCA. Assuming data $\matA$ to be properly centered, we can construct a sparse and unbiased estimate $\tilde{\matA}$ by  sampling elements from $\matA$ according to our $(\ell_1,\ell_2)$-hybrid probabilities. We then compute the principal components (PCs) of this sparse data $\tilde{\matA}$, and project original data $\matA$ onto the space spanned by these approximate PCs. Clearly, computation of PCA of sparse $\tilde{\matA}$ via SVD is computationally faster than that of PCA of $\matA$. We propose provable Algorithm 4 to compute such fast approximation of PCs.
Experimental results on synthetic and real data show promising quality of such approximation of PCA, along with faster run time.
\\ 
\\
\textit{Spectral Reconstruction of Partially-observed Data:}
Consider a situation where we do not have access to all the elements of the original matrix $\matA$. We can ask for only a small number $s$ of elements from $\matA$, and we need to predict the elements of $\matA_k$ based on these $s$ elements. We can sample (and rescale) $s$ elements of $\matA$ based on our $(\ell_1,\ell_2)$-hybrid probabilities to construct a sparse and unbiased estimator $\tilde{\matA}$ of $\matA$.   
Computation of $\tilde{\matA}_k$ via SVD can be interpreted as a spectral reconstruction of $\matA_k$ from $s$ entries of $\matA$. 
We can compute $\tilde{\matA}_k$ using Algorithm 4, and can prove the quality of $\tilde{\matA}_k$ as a surrogate for ${\matA}_k$ in Frobenius norm. Also, we can bound the quality of projection of sparsified data onto the approximate PCs. 
%For a given tolerance $\varepsilon > 0$ of such reconstruction accuracy, we like to ask for as few samples as possible from %$\matA$. That is,
%$$
%\min \quad s = \ZNorm{\tilde{\matA}}, \text{ s.t. } \TNorm{\matA - \tilde{\matA}_k} \leq \TNorm{\matA - {\matA}_k} + \varepsilon
%, \quad \ZNorm{\tilde{\matA}} \leq s
%$$
%Elements sampled according to our optimal $(\ell_1,\ell_2)$-hybrid sampling is guaranteed to require strictly smaller sample size than %$\ell_1$ or $\ell_2$ sampling for such task.
%\\  
%\\
%$\bullet$ faster approx of PCA: $\tilde{\matV}\tilde{\matV}^T\matX$ using entire $\matX$.  \\
%$\bullet$ faster approx of PCA: $\tilde{\matV}\tilde{\matV}^T\tilde{\matX}$ using sparse $\matX$.  \\
%where $\tilde{\matV}$ computed from $\tilde{\matX}$ via sparse SVD.\\
%$\bullet$ faster approx sparse PCA: $\hat{\matV}\hat{\matV}^T\matX$ using entire $\matX$.  \\
%$\bullet$ faster approx sparse PCA: $\hat{\matV}\hat{\matV}^T\tilde{\matX}$ using sparse $\matX$.  \\
%where $\hat{\matV}$ is sparsified $\tilde{\matV}$.\\
%$\hat{\matV}\hat{\matV}^T$ is NOT a projection, but $\tilde{\matV}\tilde{\matV}^T$ is.\\
%\\
%$\bullet$ sparsification comparison with leverage score sampling sampling.\\
%
}

%=================================
\subsection{Prior work}
%=================================
%
%, namely setting
%
%$$p_{ij} = \frac{\matA_{ij}^2}{\sum_{i,j=1}^{m,n} \matA_{ij}^2} = \frac{\matA_{ij}^2}{\FNormS{\matA}}.$$
%$p_{ij} = {\matA_{ij}^2}/{\FNormS{\matA}}.$
%
\cite{AM01, AM07} pioneered the idea of $\ell_2$ sampling for element-wise sparsification.
However, $\ell_2$ sampling on its own is not enough for provably accurate bounds for $\TsNorm{\matA - \tilde{\matA}}$. As a matter of fact~\cite{AM01,AM07} observed that ``small'' entries need to be sampled with probabilities that depend on their absolute values only, thus also introducing the notion of $\ell_1$ sampling. The underlying reason for the need of $\ell_1$ sampling is the fact that if a small element is sampled and rescaled using $\ell_2$ sampling, this would result in a huge entry in $\tilde{\matA}$ (because of the rescaling). As a result, the variance of $\ell_2$ sampling is quite high, resulting in poor theoretical and experimental behavior. $\ell_1$ sampling of small entries rectifies this issue by reducing the variance of the overall approach.

\cite{AHK06} proposed a sparsification algorithm that deterministically keeps large entries, i.e., entries of $\matA$ such that $\abs{\matA_{ij}} \geq \epsilon/\sqrt{n}$ and randomly rounds the remaining entries using $\ell_1$ sampling. Formally, entries of $\matA$ that are smaller than $\epsilon \sqrt{n}$ are set to $\text{sign}\left(\matA_{ij}\right)\epsilon/\sqrt{n}$ with probability $p_{ij} = \sqrt{n}\abs{\matA_{ij}}/\epsilon$ and to zero otherwise. They used an $\epsilon$-net argument to show that  $\TsNorm{\matA-\tilde{\matA}}$ was bounded with high probability. ~\cite{DZ11} bypassed the need for $\ell_1$ sampling by zeroing-out the small entries of $\matA$ (e.g., all entries such that $\abs{\matA_{ij}} < \epsilon/2n$ for a matrix $\matA \in \mathbb{R}^{n \times n}$) and then use $\ell_2$ sampling on the remaining entries in order to sparsify the matrix. This simple modification improves~\cite{AM07} and~\cite{AHK06}, and comes with an elegant proof using the matrix-Bernstein inequality of~\cite{Recht09}. %
Note that all these approaches need truncation of small entries.
%
%Note that all these approaches need \textit{a priori} access to the accuracy parameter $\epsilon>0$ as part of the input to the %algorithm in order to draw a line between large and small elements. 
Recently,~\cite{AKL13} showed that $\ell_1$ sampling in isolation could be done without any truncation, and argued that (under certain assumptions) $\ell_1$ sampling would be better than $\ell_2$ sampling, even using the truncation. Their proof is also based on the  matrix-valued Bernstein inequality of~\cite{Recht09}.
%
%=============================
%=================================
\subsection{Our Contributions}
%=================================
%
We introduce an intuitive hybrid approach to element-wise matrix sparsification, by combining $\ell_1$ and $\ell_2$ sampling. We propose to use sampling probabilities of the form
\begin{equation}\label{eqn:probability}
p_{ij} = \alpha \cdot  \frac{\abs{\matA_{ij}}}{\ONorm{\matA}}+ (1-\alpha)\frac{\matA_{ij}^2}{\FNormS{\matA}}, \qquad \alpha \in (0,1]
\end{equation}
for all $i,j$ \footnote{combining $\ell_1$ and $\ell_2$ probabilities to avoid zeroing out step of $\ell_2$ sampling has recently been observed by \cite{KD14}.}. %Here $\ONorm{\matA} = \sum_{i,j=1}^{m,n} \abs{\matA_{ij}}$.
We essentially retain the good properties of $\ell_2$ sampling that bias us towards data elements in the presence of small noise, while \textit{regularizing} smaller entries using $\ell_1$ sampling. The proof of the quality-of-approximation result of Algorithm~1 (i.e. Theorem~\ref{lem:elementsampling}) uses
the matrix-Bernstein Lemma \ref{theorem:Recht}. 
We summarize the main contributions below:

\remove{$\bullet$ Similar to~\cite{AKL13}, we do not need any zeroing out or truncation step \footnote{combining $\ell_1$ and $\ell_2$ probabilities to avoid zeroing out step of $\ell_2$ sampling has recently been observed by \cite{KD14}.} for element-wise sparsification.}

$\bullet$  We give a parameterized sampling distribution in 
the variable $\alpha \in (0, 1]$ that controls the balance between $\ell_2$ sampling and $\ell_1$ regularization. 
This greater flexibility allows us to achieve greater
accuracy.

$\bullet$ We derive the optimal hybrid-$(\ell_1,\ell_2)$ distribution, 
using Lemma \ref{theorem:Recht} for arbitrary $\matA$, by computing the optimal
parameter $\alpha^*$ which produces the desired accuracy with smallest
sample size according to our theoretical bound.

Our result generalizes the existing results because
setting $\alpha=1$ in our bounds reproduces the result of \cite{AKL13} 
who claim that $\ell_1$ sampling is almost always better
than $\ell_2$ sampling. Our results show that $\alpha^*<1$ which means that
the hybrid approach is best.

$\bullet$ We give a provable algorithm (Algorithm \ref{alg:alg_one_pass}) to implement hybrid-$(\ell_1,\ell_2)$ sampling without knowing $\alpha$ \textit{a priori}, i.e., we need not `fix' the distribution using some predetermined value of $\alpha$ at the beginning of the sampling process. We can set $\alpha$ at a later stage, yet we can realize hybrid-$(\ell_1,\ell_2)$ sampling. We use Algorithm \ref{alg:alg_one_pass} to propose a pass-efficient element-wise sampling model using only one pass over the elements of the data $\matA$, using $O(s)$ memory. Moreover, Algorithm \ref{alg:alg_iterative_alpha} gives us a heuristic to estimate $\alpha^*$ in one-pass over the data using $O(s)$ memory.

$\bullet$ Finally, we propose the Algorithm \ref{alg:alg_approx_PCA} which provably recovers PCA by constructing a sparse unbiased estimator of (centered) data using our optimal hybrid-$(\ell_1,\ell_2)$ sampling.

Experimental results suggest that our optimal hybrid distribution (using $\alpha^*$) requires strictly smaller sample size than $\ell_1$ and $\ell_2$ sampling (with or without truncation) to solve (\ref{eqn:sparsification}). Also, we achieve significant speed up of PCA on sparsified synthetic and real data while maintaining high quality approximation. 
%\\
%\\
\subsubsection{A Motivating Example for Hybrid-$(\ell_1,\ell_2)$ Sampling}
 
\begin{figure}[!h]
\centering
	\includegraphics[width=0.25\textwidth,width=5.4cm,height=3.5cm]{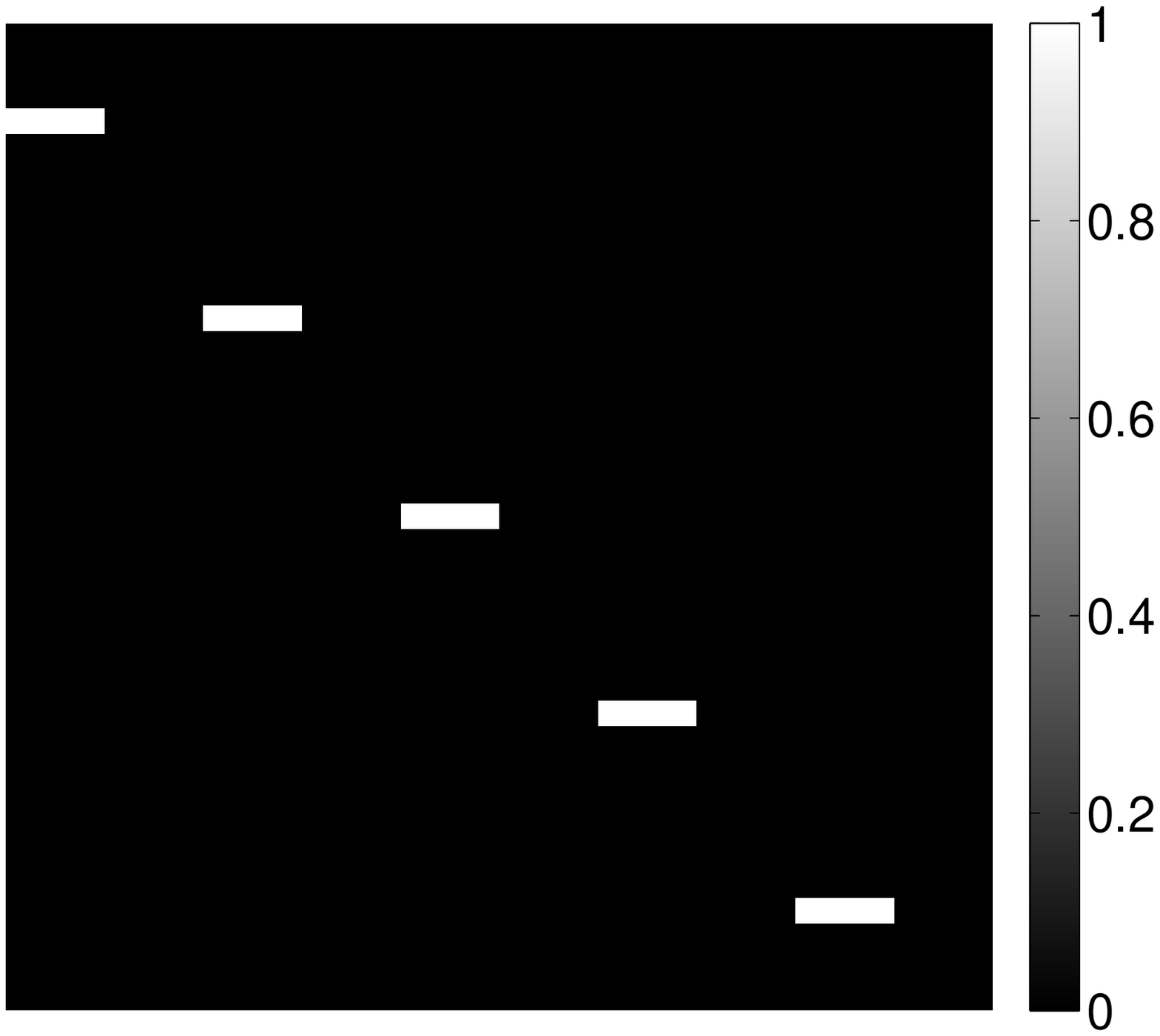}
	\label{fig:data_im}
	\qquad  
	\includegraphics[width=0.25\textwidth,width=5.4cm,height=3.5cm]{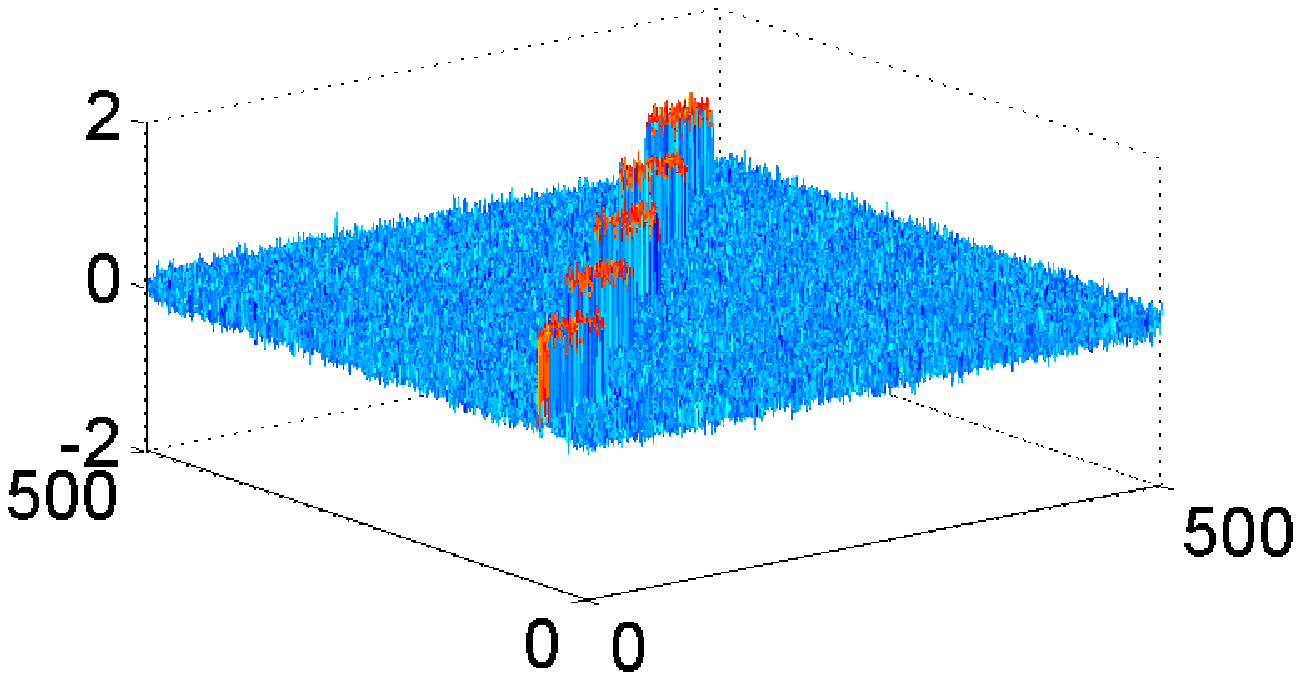}
\caption{(\textit{left}) Synthetic noiseless $500 \times 500$ binary data $\matD$; (\textit{right}) mesh view of noisy data $\matA_{0.1}$.} 
\label{fig:1}
\end{figure}

\begin{figure}[!h]
\centering
\begin{subfigure}[b]{0.3\textwidth}
	\includegraphics[height=3.2cm,width=5.2cm]{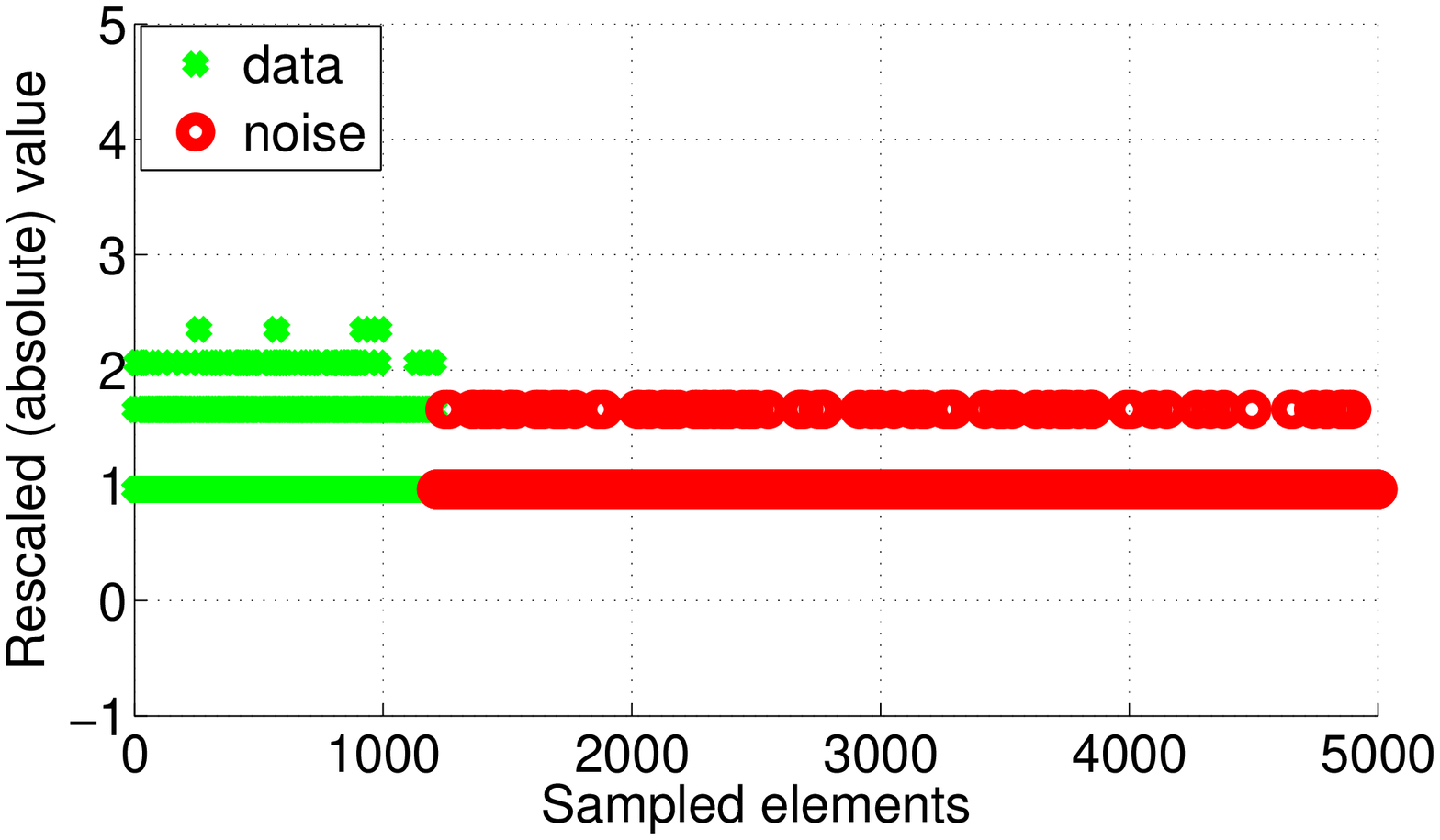}
	\caption{$\ell_1$}
\end{subfigure}
\begin{subfigure}[b]{0.3\textwidth}
	\includegraphics[height=3.2cm,width=5.2cm]{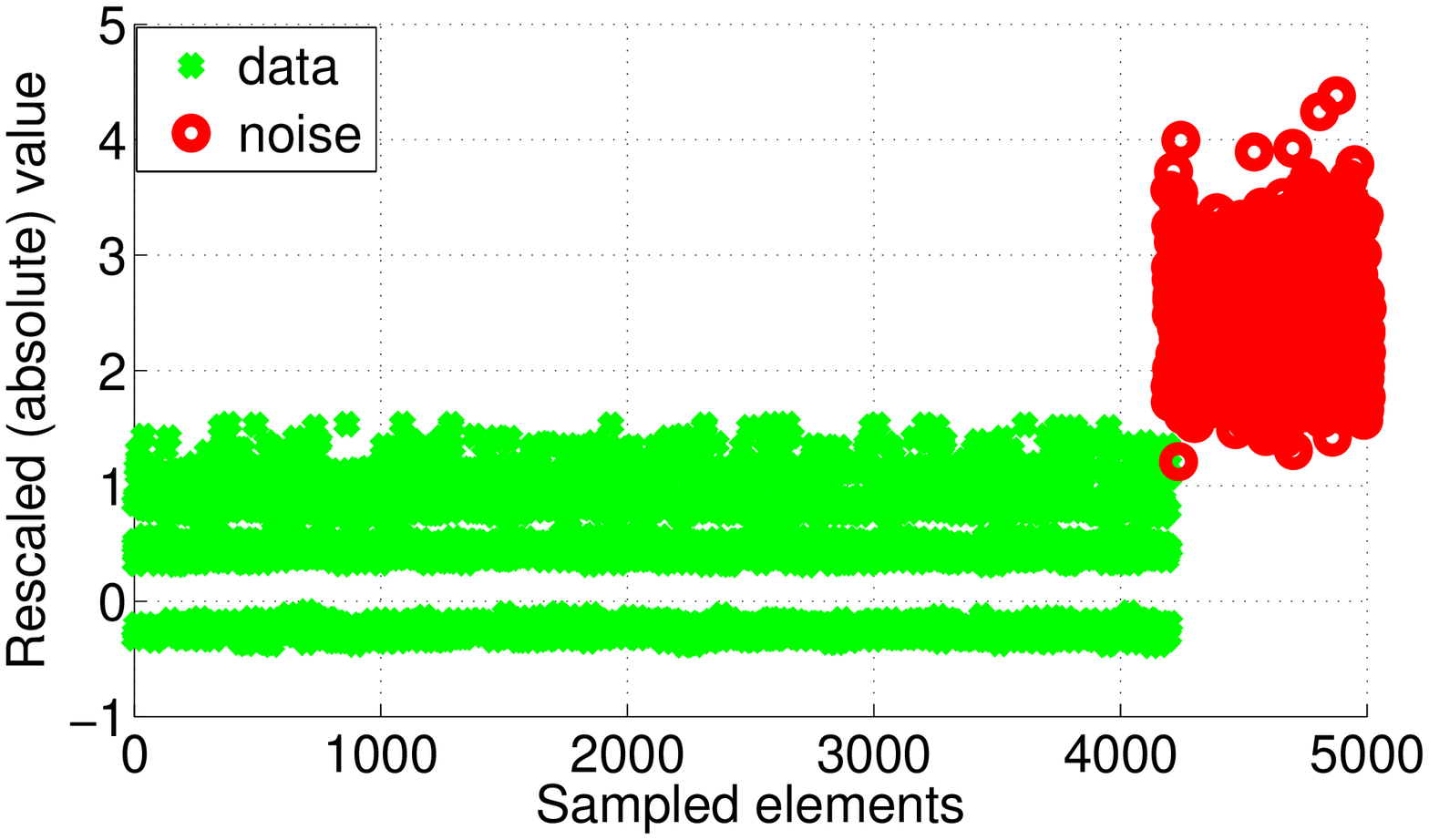}
	\caption{$\ell_2$}
\end{subfigure}
\begin{subfigure}[b]{0.3\textwidth}
	\includegraphics[height=3.2cm,width=5.2cm]{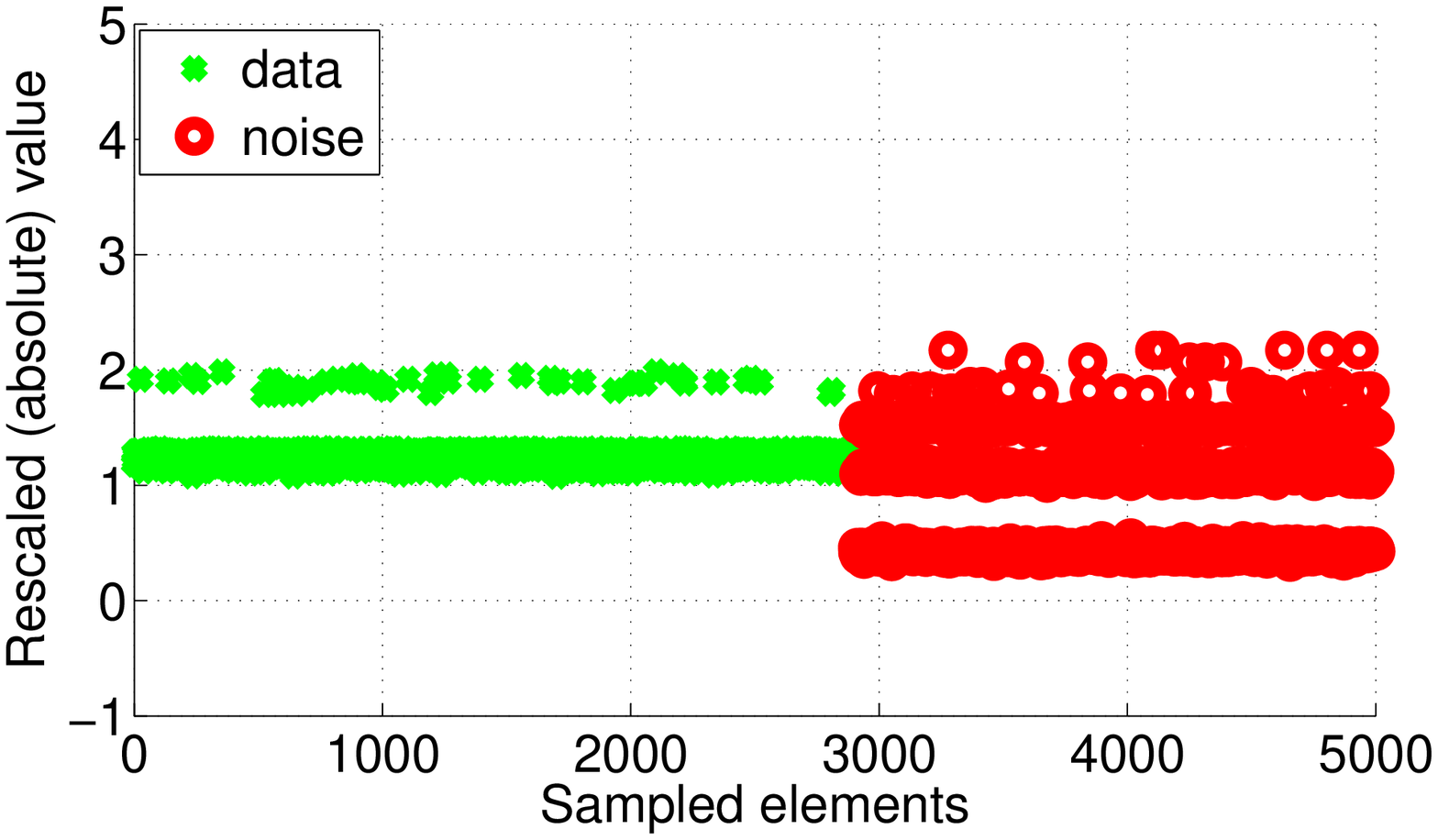}
	\caption{Hybrid-$(\ell_1,\ell_2)$}
\end{subfigure}
\caption{Elements of sparse sketch $\tilde{\matA}$ produced from $\matA_{0.1}$ via (a) $\ell_1$ sampling, (b) $\ell_2$ sampling, and (c) hybrid-$(\ell_1,\ell_2)$ sampling with $\alpha=0.7$. 
%In each plot, $x$-axis is the number of sampled indices $s= 5000$, including both data and noise. 
The $y$-axis plots the rescaled absolute values (in {$\ln$ scale}) of $\tilde{\matA}$ corresponding to the sampled indices. $\ell_1$ sampling produces elements with controlled variance but it mostly samples noise, whereas $\ell_2$ samples a lot of data although producing large variance of rescaled elements. Hybrid-$(\ell_1,\ell_2)$ sampling uses $\ell_1$ as a regularizer while sampling a fairly large number of data that helps to preserve the structure of original data.}
\label{fig:illustrate}
\end{figure}

The main motivation for introducing the idea of hybrid-$(\ell_1,\ell_2)$ 
sampling on elements of $\matA$ comes from achieving a tighter bound on $s$
 using a simple and intuitive probability distribution on elements of $\matA$. 
For this, we observe certain good properties of both $\ell_1$ and $\ell_2$ sampling for sparsification of noisy data (in practice, we experience data that are noisy, and it is perhaps impossible to separate ``true'' data from noise). 
We illustrate the behavior of $\ell_1$ and $\ell_2$ sampling on noisy data using the following synthetic example. We construct a  $500 \times 500$ binary data $\matD$ (Figure \ref{fig:1}), and then perturb it by a random Gaussian matrix  $\textbf{N}$ whose elements $\textbf{N}_{ij}$ follow Gaussian distribution with mean zero and standard deviation $0.1$. 
%=====================
%=================
We denote this perturbed data matrix by $\matA_{0.1}$. First, we note that $\ell_1$ and $\ell_2$ sampling work \textit{identically} on {binary data} $\matD$. However, Figure \ref{fig:illustrate} depicts the change in behavior of $\ell_1$ and $\ell_2$ sampling sparsifying $\matA_{0.1}$.
 Data elements and noise in $\matA_{0.1}$ are the elements with non-zero and zero values in $\matD$, respectively. We sample $s=5000$ indices in i.i.d. trials according to $\ell_1$ and $\ell_2$ probabilities separately to produce sparse sketch $\tilde{\matA}$. Figure \ref{fig:illustrate} shows that elements of $\tilde{\matA}$, produced by $\ell_1$ sampling, have controlled variance but most of them are noise. On the other hand, $\ell_2$ sampling is biased towards data elements, although small number of sampled noisy elements create large variance due to rescaling. Our hybrid-$(\ell_1,\ell_2)$ sampling benefits from this bias of $\ell_2$ towards data elements, as well as, regularization properties of $\ell_1$.
%
% ============================================
%
%
%=====================
%Thus, our choice of $\alpha^*$ always performs better than other extreme choices of sampling.

We parameterize our distribution using the variable $\alpha \in (0, 1]$ that controls the balance between $\ell_2$ sampling and $\ell_1$ regularization. We derive an expression to compute $\alpha^*$,  the optimal $\alpha$, corresponding to the smallest sample size that we need in order to achieve a given accuracy $\epsilon$ in (\ref{eqn:sparsification}). Setting $\alpha=1$, we reproduce the result of~\cite{AKL13}. However, $\alpha^*$ may be smaller than 1, and the bound on sample size $s$, using $\alpha^*$, is guaranteed to be tighter than that of~\cite{AKL13}.
%
%
%===================================
\section{Main Result}
%===================================
%
%\subsection{Notation and Matrix Metrics}\label{sec:notation}
%
%
%\subsection{Our main theorem}

\begin{figure}[!h]
\centering
\includegraphics[height=3.5cm,width=5.6cm]{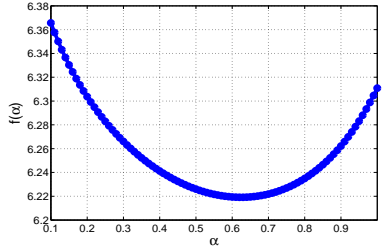}
\caption{Plot of $f(\alpha)$ in eqn (\ref{eqn:opt_alpha}) for data $\matA_{0.1}$. We use $\epsilon = 0.05$ and $\delta=0.1$. $x$-axis plots $\alpha$ and $y$-axis is in log$_{10}$ scale. For this data, $\alpha^* \approx 0.6$.}
\label{fig:opt_alpha}
\end{figure}

We present the quality-of-approximation result of our main algorithm (Algorithm 1). 
We define the sampling operator $\mathcal{S}_{\Omega}: \mathbb{R}^{m \times n} \rightarrow \mathbb{R}^{m \times n}$ in (\ref{eqn:sampling_operator}) that extracts elements from a given matrix $\matA \in \mathbb{R}^{m \times n}$. Let $\Omega$ be a multi-set of sampled indices $(i_t,j_t)$, for $t=1,...,s$. Then,
\begin{eqnarray}\label{eqn:sampling_operator}
\mathcal{S}_{\Omega}\left(\matA\right) = \frac{1}{s}\sum_{t=1}^s \frac{\matA_{i_t j_t}}{p_{i_t j_t}} \e_{i_t} \e_{j_t}^T, 
\quad (i_t,j_t) \in \Omega 
\end{eqnarray}
Algorithm~1 randomly samples (in i.i.d. trials) $s$ elements of a given matrix $\matA$, according to a probability distribution $\{p_{ij}\}_{i,j=1}^{m,n}$ over the elements of $\matA$. Let the $p_{ij}$'s be as in eqn.~(\ref{eqn:probability}). Then, we can prove the following theorem.
\begin{theorem}\label{lem:elementsampling}
Let $\matA \in \mathbb{R}^{m \times n}$ and  let $\epsilon > 0$ be an accuracy parameter. Let $\mathcal{S}_{\Omega}$ be the sampling operator defined in (\ref{eqn:sampling_operator}), and assume that the multi-set $\Omega$ is generated using sampling probabilities $\left\{p_{ij}\right\}_{i,j=1}^{m,n}$ as in (\ref{eqn:probability}).
%
%\begin{equation*}
%
%p_{ij} = \alpha \cdot  \frac{\abs{\matA_{ij}}}{\ONorm{\matA}}+ (1-\alpha)\frac{\matA_{ij}^2}{\FNormS{\matA}}, \qquad \alpha \in (0,1]
%
%\end{equation*}
%
%for all $i,j$. 
Then, with probability at least $1 - \delta$,
\begin{eqnarray}\label{eqn:stable_rank_bound}
\TNorm{\mathcal{S}_{\Omega}(\matA) - \matA} \leq \epsilon \TNorm{\matA},
\end{eqnarray}
if
\begin{eqnarray}\label{eqn:s_alpha}
s \geq \frac{2}{\epsilon^2\TNorm{\matA}^2}\left( \rho^2(\alpha) + \gamma(\alpha)\epsilon\TNorm{\matA}/3\right) \ln\left(\frac{m+n}{\delta}\right)
\end{eqnarray}
where,  
$$
\xi_{ij} = {\FNormS{\matA}}/\left (\frac{\alpha \cdot  \FNormS{\matA}}{\abs{\matA_{ij}}\cdot \ONorm{\matA}}+ (1-\alpha)\right ),
\text{ for } \matA_{ij}\neq 0,
$$
%$$\rho^2(\alpha) = \max\left\{\max_i \sum_{j=1}^{n}\frac{\FNormS{\matA}}{\frac{\alpha \cdot  \FNormS{\matA}}{\abs{\matA_{ij}}\cdot \ONorm{\matA}}+ (1-\alpha)},\max_j \sum_{i=1}^{m}\frac{\FNormS{\matA}}{\frac{\alpha \cdot  \FNormS{\matA}}{\abs{\matA_{ij}}\cdot \ONorm{\matA}}+ (1-\alpha)}\right\} - \sigma_{min}^2(\matA),$$
%
$$\rho^2(\alpha) = \max\left\{\max_i \sum_{j=1}^{n}\xi_{ij},\max_j \sum_{i=1}^{m}\xi_{ij}\right\} - \sigma_{min}^2(\matA),$$
$$\gamma(\alpha) = \max_{\stackrel{i,j:}{\matA_{ij}\neq 0}}\left\{\frac{\ONorm{\matA}}{\alpha + (1-\alpha)\frac{\ONorm{\matA}\cdot \abs{\matA_{ij}}}{\FNormS{\matA}}}\right\} + \TNorm{\matA},$$
$\sigma_{min}(\matA)$ is the smallest singular value of $\matA$.
Moreover, we can find $\alpha^*$ (optimal $\alpha$ corresponding to the smallest $s$) and $s^*$ (the smallest $s$), by solving the following optimization problem in (\ref{eqn:opt_alpha}):
\begin{eqnarray}\label{eqn:opt_alpha}
%\alpha^* = \min_{\alpha \in (0,1]}\left\{ \rho^2(\alpha) + \gamma(\alpha)\epsilon\TNorm{\matA}/3\right\}
\alpha^* = \min_{\alpha \in (0,1]}f(\alpha), \quad  
f(\alpha) = \rho^2(\alpha) + \gamma(\alpha)\epsilon\TNorm{\matA}/3,
\end{eqnarray}
\begin{eqnarray}\label{eqn:opt_s}
s^* = \frac{2}{\epsilon^{2}\TNorm{\matA}^2}\left( \rho^2(\alpha^*) + \gamma(\alpha^*)\frac{\epsilon\TNorm{\matA}}{3}\right) \ln\left(\frac{m+n}{\delta}\right)
\end{eqnarray}
%
%
%\max_{ij}\left\{\left(\frac{\alpha}{\ONorm{\matA}} + \frac{(1-\alpha)\cdot \abs{\matA_{ij}}}{\FNormS{\matA}}\right)^{-1}\right\} + \TNorm{\matA}
%$$
% s \geq \frac{2}{\epsilon^2\TNorm{\matA}^2}\left(\max_i \left\{\sum_{j=1}^{n}\frac{\FNormS{\matA}}{\frac{\alpha \cdot  %\FNormS{\matA}}{\abs{\matA_{ij}}\cdot \ONorm{\matA}}+ (1-\alpha)}\right\} - \sigma_{min}^2(\matA) + %\left(\frac{\ONorm{\matA}}{\alpha} + \TNorm{\matA}\right)\frac{\epsilon\cdot\TNorm{\matA}}{3 }\right)\ln\left(\frac{m+n}%{\delta}\right).
%
%$$
%
%Further, if 
%$\varepsilon\leq 3\TNorm{\matA}^{-1} \left(\frac{\ONorm{\matA}}{\alpha} + \TNorm{\matA}\right)^{-1}\left(\max_i %\left\{\sum_{j=1}^{n}\frac{\FNormS{\matA}}{\frac{\alpha \cdot  \FNormS{\matA}}{\abs{\matA_{ij}}\cdot \ONorm{\matA}}+ %(1-\alpha)}\right\} - \sigma_{min}^2(\matA)\right)$, and
%$$
% s \geq \frac{4}{\varepsilon^2}\text{sr}(\matA)\left(\max_i \left\{\sum_{j=1}^{n}\frac{1}{\frac{\alpha \cdot  
%\FNormS{\matA}}{\abs{\matA_{ij}}\cdot \ONorm{\matA}}+ (1-\alpha)}\right\} - \frac{\sigma_{min}^2(\matA)}{\FNormS{\matA}} %\right)\ln\left(\frac{m+n}{\delta}\right),
%
%$$
%then, with probability at least $1 - \delta$,
%
%\begin{eqnarray}\label{eqn:stable_rank_bound}
%\TNorm{\matA - \mathcal{S}_{\Omega}(\matA)} \leq \varepsilon\TNorm{\matA}.
%\end{eqnarray}
%
\end{theorem}
%
%We essentially parameterize the key quantities $\rho^2$ and $\gamma$ of Matrix-Bernstein inequality in Lemma %\ref{theorem:Recht} using the distribution parameter $\alpha$. This gives us a flexibility to express the sample size as a function %of $\alpha$ in (\ref{eqn:s_alpha}). Naturally, we want to solve (\ref{eqn:opt_alpha}) to find the $\alpha$ that gives us the %smallest value of $s$.
%
The functional form in (\ref{eqn:s_alpha}) comes from the Matrix-Bernstein inequality in Lemma \ref{theorem:Recht}, with $\rho^2$ and $\gamma$ being functions of $\matA$ and $\alpha$. This gives us a flexibility to optimize the sample size with respect to $\alpha$ in (\ref{eqn:s_alpha}), which is how we get the optimal $\alpha^*$.   
For a given matrix $\matA$, we can easily compute $\rho^2(\alpha)$ and $\gamma(\alpha)$ for various values of $\alpha$. Given an  accuracy $\epsilon$ and failure probability $\delta$, we can compute $\alpha^*$ corresponding to the tightest bound on $s$. 
Note that, for $\alpha=1$ we reproduce the results of \cite{AKL13} (which was expressed using various matrix metrics). 
However, $\alpha^*$ may be smaller than 1, and is guaranteed to produce tighter $s$ comparing to extreme choices of $\alpha$ (e.g. $\alpha = 1$ for $\ell_1$ sampling). We illustrate this by the plot in Figure \ref{fig:opt_alpha}.
%(see Section \ref{experiments} for description of data). 
%---------------------
\begin{algorithm}[t]
\centerline{
\caption{Element-wise Matrix Sparsification }\label{alg:alg1}
}
\begin{algorithmic}[1]
%--------------------------------------------------
\STATE \textbf{Input:} $\matA \in \mathbb{R}^{m \times n}$, accuracy parameter $\epsilon > 0$.
%\STATE \textbf{Input:} $\matA \in \mathbb{R}^{m \times n}$, $\left\{p_{ij}\right\}_{i,j=1}^{m,n}$ such that $p_{ij} \geq 0$ (for all %$i,j$) and $\sum_{i,j=1}^{m,n} p_{ij}=1$, integer $s > 0$.
%--------------------------------------------------
\STATE \textbf{Set} $s$ as in eq. (\ref{eqn:opt_s}).
%\STATE \textbf{For} $t = 1\ldots s$ (i.i.d. trials with replacement) \textbf{randomly sample} pairs of indices $(i_t, j_t) \in \{1\ldots m\}\times \{1\ldots n\}$ with
%
\STATE \textbf{For} $t = 1\ldots s$ (i.i.d. trials with replacement) \textbf{randomly sample} pairs of indices $(i_t, j_t) \in [m]\times [n]$ with
$\mathbb{P}\left[ (i_t, j_t) =  (i,j)\right]\ =\ p_{ij}$, where $p_{ij}$ are as in (\ref{eqn:probability}), using $\alpha$ as in (\ref{eqn:opt_alpha}).
%
%\STATE \textbf{Output:} set of sampled pairs of indices $\Omega = \left\{\left(i_t,j_t\right),t=1\ldots s\right\}$, along with rescaled values $\frac{\matA_{i_t j_t}}{s\cdot p_{i_t j_t}}$.
%\STATE \textbf{Output:} $\Omega = \left\{\left(i_t,j_t\right),t=1\ldots s\right\}$, along with rescaled values $\frac{\matA_{i_t j_t}}%{s\cdot p_{i_t j_t}}$.
%
%\STATE \textbf{Sampling operator:} $\mathcal{S}_{\Omega}: \mathbb{R}^{m \times n}\rightarrow \mathbb{R}^{m \times n}$ with
\STATE \textbf{Output}(sparse): 
$\mathcal{S}_{\Omega}\left(\matA\right) = \frac{1}{s}\sum_{t=1}^s \frac{\matA_{i_t j_t}}{p_{i_t j_t}} \e_{i_t} \e_{j_t}^T.$
\end{algorithmic}
\end{algorithm} 

%----------------------
%
We give a proof of Theorem \ref{lem:elementsampling} in Section \ref{sxn:appendix:primitive2}.
%Rest of the paper contains a proof of Theorem \ref{lem:elementsampling}, followed by, discussion on pass-efficient implementation of $(\ell_1,\ell_2)$-hybrid sampling, and finally experiments on various data.
%
%Now, we present the main algorithm.%
%
%
\subsection{Proof of Theorem~\ref{lem:elementsampling}}\label{sxn:appendix:primitive2}
In this section we provide a proof of Theorem~\ref{lem:elementsampling} following the proof outline of~\cite{DZ11, AKL13}. 
We use the following non-commutative matrix-valued Bernstein bound of \cite{Recht09} as our main tool to prove Theorem \ref{lem:elementsampling}. Using our notation we rephrase the matrix Bernstein bound.
\begin{lemma}\label{theorem:Recht}
[Theorem 3.2 of \cite{Recht09}] Let $\matM_1, \matM_2, ..., \matM_s$ be independent, zero-mean random matrices in $\mathbb{R}^{m \times n}$. Suppose $$\max_{t\in [s]} \left\{\TNorm{\mathbb{E}(\matM_t\matM_t^T)}, \TNorm{\mathbb{E}(\matM_t^T\matM_t)} \right\} \leq \rho^2$$ and $\TNorm{\matM_t} \leq \gamma$ for all $t \in [s]$. Then, for any $\epsilon > 0$,
$$\TNorm{\frac{1}{s}\sum_{t=1}^{s}\matM_t} \leq \epsilon$$
holds, subject to a failure probability at most
$$\left ( m+n\right )\exp\left ( \frac{-s\epsilon^2/2}{\rho^2 + \gamma\epsilon /3} \right ).$$
\end{lemma}
For all $t \in [s]$ we define the matrix $\matM_t \in \mathbb{R}^{m \times n}$ as follows:
$$\matM_t = \frac{\matA_{i_tj_t}}{p_{i_tj_t}}\e_{i_t}\e_{j_t}^T - \matA.$$
%\begin{eqnarray}\label{eqn:Mt}
%\matM_t = \frac{\matA_{i_tj_t}}{p_{i_tj_t}}\e_{i_t}\e_{j_t}^T - \matA.
%\end{eqnarray}
%
It now follows that
$$\frac{1}{s}\sum_{t=1}^{s}\matM_t = \frac{1}{s}\sum_{t=1}^{s}\left[ \frac{\matA_{i_tj_t}}{p_{i_tj_t}}\e_{i_t}\e_{j_t}^T - \matA \right ] = S_{\Omega}(\matA) - \matA.$$ 
We can bound $\TNorm{\matM_t}$ for all $t \in [s]$. We define the following quantity:
\begin{eqnarray}\label{eqn:lambda}
\lambda = \frac{\ONorm{\matA}\cdot \abs{\matA_{ij}}}{\FNormS{\matA}}, \text{ for } \matA_{ij} \neq 0
\end{eqnarray}
\begin{lemma}\label{lemma:boundMt}
%Using our notation, $\TNorm{\matM_t} \leq \frac{\ONorm{\matA}}{\alpha}+\TNorm{\matA}$ for all $t \in [s]$.
%
Using our notation, and using probabilities of the form (\ref{eqn:probability}), for all $t \in [s]$, 
$$
\TNorm{\matM_t} \leq \max_{\stackrel{i,j:}{\matA_{ij}\neq 0}}\frac{\ONorm{\matA}}{\alpha+(1-\alpha)\lambda}+\TNorm{\matA}.
$$
\end{lemma}
\begin{Proof}
Using probabilities of the form (\ref{eqn:probability}), and because $\matA_{ij}=0$ is never sampled,
\begin{eqnarray*}
 \TNorm{\matM_t} =  \TNorm{\frac{\matA_{i_tj_t}}{p_{i_tj_t}}\e_{i_t}\e_{j_t}^T - \matA}
\leq \max_{\stackrel{i,j:}{\matA_{ij}\neq 0}}\left\{\left(\frac{\alpha}{\ONorm{\matA}} + \frac{(1-\alpha)\cdot \abs{\matA_{ij}}}{\FNormS{\matA}}\right)^{-1}\right\} + \TNorm{\matA}
\end{eqnarray*}
%
%In the above we used $0 < \alpha\leq 1$. %$\TNorm{\matA} \leq \ONorm{\matA}$.\\
%\\
Using (\ref{eqn:lambda}), we obtain the bound.
%Thus, we get a new bound for Lemma 2 of \cite{DZ11}, bypassing the need for a truncation step.
%
\end{Proof}
\\
\noindent Next we bound the spectral norm of the expectation of $\matM_t\matM_t^T$.
% The spectral norm of the expectation of $\matM_t^T\matM_t$ can be bounded using a similar analysis.
%
\begin{lemma}\label{lemma:boundvar} Using our notation, and using probabilities of the form (\ref{eqn:probability}), for all $t \in [s]$, 
$$
\TNorm{\mathbb{E}(\matM_t\matM_t^T)} \leq \FNormS{\matA} \beta_1 - \sigma_{min}^2(\matA),$$ where, $$\beta_1 = \max_i \sum_{j=1}^{n}\left(\frac{\alpha \cdot  \FNormS{\matA}}{\abs{\matA_{ij}}\cdot \ONorm{\matA}}+ (1-\alpha)\right)^{-1},
\text{ for } \matA_{ij}\neq 0.
$$
\end{lemma}
\begin{Proof}
Recall that $\matA = \sum_{i,j=1}^{m,n}\matA_{ij}\e_{i}\e_{j}^T$ and $\matM_t = \frac{\matA_{i_tj_t}}{p_{i_tj_t}}\e_{i_t}\e_{j_t}^T - \matA$ to derive
\begin{eqnarray*}
\mathbb{E}[\matM_{t}\matM_{t}^T]
&=& \sum_{i,j=1}^{m,n} p_{ij}\left(\frac{\matA_{ij}}{p_{ij}}\e_{i}\e_{j}^T - \matA\right) \left(\frac{\matA_{ij}}{p_{ij}}\e_{j}\e_{i}^T - \matA^T\right)\\
&=& \sum_{i,j=1}^{m,n}\left(\frac{\matA_{ij}^2}{p_{ij}} \e_{i}\e_{i}^T\right) - \matA \matA^T.
\end{eqnarray*}
\noindent Sampling according to probabilities of eqn.~(\ref{eqn:probability}), and because $\matA_{ij}=0$ is never sampled, we get,  for $\matA_{ij}\neq 0$,
\begin{eqnarray*}
\sum_{i,j=1}^{m,n}\frac{\matA_{ij}^2}{p_{ij}}
&=& \FNormS{\matA}\sum_{i,j=1}^{m,n}\left(\frac{\alpha \cdot  \FNormS{\matA}}{\abs{\matA_{ij}}\cdot \ONorm{\matA}}+ (1-\alpha)\right)^{-1}, \\
%
%$\leq \FNormS{\matA}\sum_{i=1}^{m}\max_i\sum_{j=1}^{n}\frac{1}{\frac{\alpha \cdot  \FNormS{\matA}}%{\abs{\matA_{ij}}\cdot \ONorm{\matA}}+ (1-\alpha)}.
%$\\
%
&\leq& \FNormS{\matA}\sum_{i=1}^{m}\max_i\sum_{j=1}^{n}\left(\frac{\alpha \cdot  \FNormS{\matA}}{\abs{\matA_{ij}}\cdot \ONorm{\matA}}+ (1-\alpha)\right)^{-1}. %\text{ for } \matA_{ij}\neq 0.
\end{eqnarray*}
Thus, 
\begin{eqnarray*}
\mathbb{E}[\matM_{t}\matM_{t}^T] 
\preceq \FNormS{\matA} \beta_1 \sum_{i=1}^{m}\e_{i}\e_{i}^T - \matA \matA^T
= \FNormS{\matA} \beta_1 \matI_{m} - \matA \matA^T.
\end{eqnarray*}
%
%$$\mathbb{E}[\matM_{t}\matM_{t}^T] = \sum_{i,j=1}^{m,n}\left(\frac{\matX_{ij}^2}{p_{ij}}\e_{i}\e_{i}^T\right) - \matX \matX^T \preceq \frac{2\FNormS{\matX}}{\beta} \sum_{i,j=1}^{m,n}\e_{i}\e_{i}^T - \matX \matX^T = \frac{2n\FNormS{\matX}}{\beta} \matI_{m} - \matX \matX^T.$$
%
%Using Weyl's inequality we get
Note that, $\FNormS{\matA} \beta_1 \matI_{m}$ is a diagonal matrix with all entries non-negative, and $\matA\matA^T$ is a postive semi-definite matrix. Therefore,\\
%
%\begin{eqnarray*}
$$\TNorm{\mathbb{E}[\matM_{t}\matM_{t}^T]} \leq \FNormS{\matA} \beta_1 - \sigma_{min}^2(\matA).$$
%\end{eqnarray*}
%
%assuming, ${\text{rs}_1(\matA)}\frac{\ONormS{\matA}}{m\alpha } \geq \TNormS{\matA}$.
%
\end{Proof}
\\
Similarly, we can obtain
$$\TNorm{\mathbb{E}[\matM_{t}^T\matM_{t}]} \leq \FNormS{\matA} \beta_2 - \sigma_{min}^2(\matA),$$ where,\\
$$
\beta_2 = \max_j \sum_{i=1}^{m}\left(\frac{\alpha \cdot  \FNormS{\matA}}{\abs{\matA_{ij}}\cdot \ONorm{\matA}}+ (1-\alpha)\right)^{-1}, \text{ for } \matA_{ij}\neq 0.
$$
We can now apply Theorem~\ref{theorem:Recht} with $$\rho^2(\alpha) =  \FNormS{\matA} \max\{\beta_1,\beta_2\} - \sigma_{min}^2(\matA)$$ and 
%$\gamma(\alpha) =  \frac{\ONorm{\matA}}{\alpha} + \TNorm{\matA}$
%
$$
\gamma(\alpha) = \frac{\ONorm{\matA}}{\alpha+(1-\alpha)\lambda}+\TNorm{\matA}
$$
  to conclude that
$\TNorm{\mathcal{S}_{\Omega}(\matA) -  \matA} \leq \varepsilon$ holds subject to a failure probability at most
%
%$\left(m+n\right)\exp\left(\frac{-s\epsilon^2/2}{ \frac{\text{rs}_1(\matA)}{m\alpha }\ONormS{\matA} + \frac{2}{3\alpha %}\epsilon\ONorm{\matA}}\right).$\\
%
$$\left(m+n\right)\exp\left({(-s\varepsilon^2/2)}/{\left(\rho^2(\alpha) + \gamma(\alpha)\varepsilon/3 \right)}\right).$$ Bounding the failure probability by $\delta$, and setting $\varepsilon = \epsilon\cdot \TNorm{\matA},$ we complete the proof.
%
%Setting the failure probability equal to $\delta$, we conclude that it suffices to set $s$ as follows:
%
%\begin{eqnarray*}
%
%$$
% s \geq \frac{2}{\epsilon^2}\left(\max_i \left\{\sum_{j=1}^{n}\frac{\FNormS{\matA}}{\frac{\alpha \cdot  \FNormS{\matA}}%{\abs{\matA_{ij}}\cdot \ONorm{\matA}}+ (1-\alpha)}\right\} - \sigma_{min}^2(\matA) + %\left(\frac{\ONorm{\matA}}{\alpha} + \TNorm{\matA}\right)\frac{\epsilon}{3 }\right)\ln\left(\frac{m+n}{\delta}\right).
%%
%$$
%%\end{eqnarray*}
%%
%Let, $\epsilon = \varepsilon\TNorm{\matA} \leq 3\left(\frac{\ONorm{\matA}}{\alpha} + \TNorm{\matA}\right)^{-1}\left(\max_i %\left\{\sum_{j=1}^{n}\frac{\FNormS{\matA}}{\frac{\alpha \cdot  \FNormS{\matA}}{\abs{\matA_{ij}}\cdot \ONorm{\matA}}+ %(1-\alpha)}\right\} - \sigma_{min}^2(\matA)\right)$. \\
%Recall, $\text{sr}(\matA) = {\FNormS{\matA}}/{\TNormS{\matA}}$. Thus, if
%$$
% s \geq \frac{4}{\varepsilon^2}\text{sr}(\matA)\left(\max_i \left\{\sum_{j=1}^{n}\frac{1}{\frac{\alpha \cdot  \FNormS{\matA}}{\abs{\matA_{ij}}\cdot \ONorm{\matA}}+ (1-\alpha)}\right\} - \frac{\sigma_{min}^2(\matA)}{\FNormS{\matA}} \right)\ln\left(\frac{m+n}{\delta}\right).
%
%$$
%then, with probability at least $1 - \delta$,
%
%$$\TNorm{\matA - \mathcal{S}_{\Omega}(\matA)} \leq \varepsilon\TNorm{\matA}.$$
%

%\input{Proofs}
%\newpage
\section{One-pass Hybrid-$(\ell_1,\ell_2)$ Sampling}
Here we discuss the implementation of $(\ell_1,\ell_2)$-hybrid sampling in one pass over the input matrix $\matA$ using $O(s)$ memory, that is, a streaming model. We know that both $\ell_1$ and $\ell_2$ sampling can be done in one pass using $O(s)$ memory (see Algorithm SELECT p. 137 of \cite{DKM06} ). 
%
%Similarly, if we know the value of $\alpha$ beforehand then we can perform the hybrid sampling in one pass using $O(s)$ memory. 
%
In our hybrid sampling, we want parameter $\alpha$ to depend on data elements, i.e., we do not want to `fix' it prior to the arrival of data stream.
% However, $\alpha$ depends on data and is unlikely to be known beforehand. 
Here we give an algorithm (Algorithm \ref{alg:alg_one_pass}) to implement a one-pass version of the hybrid sampling \textit{without knowing $\alpha$ a priori}.
\begin{algorithm}[t]
\centerline{
\caption{One-pass hybrid-$(\ell_1,\ell_2)$ sampling}\label{alg:alg_one_pass}
}
\begin{algorithmic}[1]
%--------------------------------------------------
\STATE \textbf{Input:} $\matA_{ij}$ for all $(i,j) \in [m]\times [n]$, arbitrarily ordered, and sample size $s$.%, and $\alpha \in [0,1]$.
%--------------------------------------------------
%\STATE Apply SELECT algorithm in parallel with $O(s)$ memory using $L_1$ probabilities to sample $s$ independent indices $(i_{t_1},j_{t_1})$ and corresponding rescaled elements $\matA_{i_{t_1}j_{t_1}}/p_{i_{t_1}j_{t_1}}$ to form a random multiset $S_1$ of triples $(i_{t_1}, j_{t_1}, \matA_{i_{t_1}j_{t_1}}/p_{i_{t_1}j_{t_1}})$, for $t_1 = 1,... , s$. 
%
\STATE Apply SELECT algorithm in parallel with $O(s)$ memory using $\ell_1$ probabilities to sample $s$ independent indices $(i_{t_1},j_{t_1})$ and corresponding elements $\matA_{i_{t_1}j_{t_1}}$ to form random multiset $S_1$ of triples $(i_{t_1}, j_{t_1}, \matA_{i_{t_1}j_{t_1}})$, for $t_1 = 1,... , s$. 
%
%\STATE Apply SELECT algorithm in parallel with $O(s)$ memory using $L_1$ probabilities to sample $s$ independent indices $(i_{t_3},j_{t_3})$ and corresponding elements $\matA_{i_{t_3}j_{t_3}}$ to form random multiset $S_3$ of triples $(i_{t_3}, j_{t_3}, \matA_{i_{t_3}j_{t_3}})$, for $t_3 = 1,... , s$. %($S_3$ is independent of $S_1$)
%
\STATE Run step 2 in parallel to form another independent multiset $S_3$ of triples $(i_{t_3}, j_{t_3}, \matA_{i_{t_3}j_{t_3}})$, for $t_3 = 1,... , s$. (This step is only for Algorithm \ref{alg:alg_iterative_alpha})
%($S_3$ is independent of $S_1$)
%
%
\STATE Apply SELECT algorithm in parallel with $O(s)$ memory using $\ell_2$ probabilities to sample $s$ independent indices $(i_{t_2},j_{t_2})$ and corresponding elements $\matA_{i_{t_2}j_{t_2}}$ to form random multiset $S_1$ of triples $(i_{t_2}, j_{t_2}, \matA_{i_{t_2}j_{t_2}})$, for $t_2 = 1,... , s$. 
%
%\STATE Apply SELECT algorithm in parallel with $O(s)$ memory using $L_2$ probabilities to sample $s$ independent indices $(i_{t_4},j_{t_4})$ and corresponding elements $\matA_{i_{t_4}j_{t_4}}$ to form random multiset $S_4$ of triples $(i_{t_4}, j_{t_4}, \matA_{i_{t_4}j_{t_4}})$, for $t_4 = 1,... , s$. 
%
\STATE Run step 4 in parallel to form another independent multiset $S_4$ of triples $(i_{t_4}, j_{t_4}, \matA_{i_{t_4}j_{t_4}})$, for $t_4 = 1,... , s$. (This step is only for Algorithm \ref{alg:alg_iterative_alpha})
\STATE Compute and store $\FNormS{\matA}$ and $\ONorm{\matA}$ in parallel.
\STATE Set the value of $\alpha \in (0,1]$ (using Algorithm \ref{alg:alg_iterative_alpha}).
%\STATE Set the value of parameter $\alpha$.
\STATE Create empty multiset of triples $S$. % \leftarrow \emptyset$.
\STATE $\matX \leftarrow \textbf{0}_{m\times n}$. 
\STATE \textbf{For} $t = 1\ldots s$ \\
\STATE \quad Generate a uniform random number $x \in [0,1]$.\\% , $x \sim \text{unif}[0,1]$.\\
\STATE \quad if $x \geq \alpha$,  $S(t) \leftarrow S_1(t)$; otherwise, $S(t) \leftarrow S_2(t)$.\\
\STATE \quad $(i_t,j_t) \leftarrow S(t,1:2)$.
\STATE \quad $p \leftarrow \alpha\cdot\frac{\abs{S(t,3)}}{\ONorm{\matA}} + (1-\alpha)\cdot\frac{\abs{S(t,3)}^2}{\FNormS{\matA}}$ 
\STATE \quad $\matX \leftarrow \matX + \frac{S(t,3)}{p\cdot s}e_{i_t}e_{j_t}^T$.
\STATE \textbf{End}
%
%
%\STATE \textbf{Output:} set of sampled pairs of indices $\Omega = \left\{\left(i_t,j_t\right),t=1\ldots s\right\}$, along with rescaled values $\frac{\matA_{i_t j_t}}{s\cdot p_{i_t j_t}}$.
\STATE \textbf{Output:} random multiset $S$, and sparse matrix $\matX$.
%
%\STATE \textbf{Sampling operator:} $\mathcal{S}_{\Omega}: \mathbb{R}^{m \times n}\rightarrow \mathbb{R}^{m \times n}$ with
%
%$\mathcal{S}_{\Omega}\left(\matA\right) = \frac{1}{s}\sum_{t=1}^s \frac{\matA_{i_t j_t}}{p_{i_t j_t}} \e_{i_t} \e_{j_t}^T.$
%
\end{algorithmic}
\end{algorithm}

We note that steps 2-5 of Algorithm \ref{alg:alg_one_pass} access the elements of $\matA$ only once, in parallel, to form independent multisets $S_1$, $S_2$, $S_3$, and $S_4$. Step 6 computes $\FNormS{\matA}$ and $\ONorm{\matA}$ in parallel in one pass over $\matA$. Subsequent steps do not need to access $\matA$ anymore. 
Interestingly, we set $\alpha$ in step 7 when the data stream is gone. Steps 10-16 sample $s$ elements from $S_1$ and $S_2$ based on the $\alpha$ in step 7, and produce sparse matrix $\matX$ based on the sampled entries in random multiset $S$.
Theorem \ref{thm:one_pass} shows that Algorithm \ref{alg:alg_one_pass} indeed samples elements from $\matA$ according to the hybrid-$(\ell_1,\ell_2)$ probabilities in eqn (\ref{eqn:probability}).
\begin{theorem}\label{thm:one_pass}
Using the notations in Algorithm 2, for $\alpha \in (0,1]$, $t = 1, ..., s$, 
$$
\prob{S(t) = (i,j,\matA_{ij})} = \alpha \cdot p_1 + (1-\alpha)\cdot p_2,
%\prob{S(t) = (i,j,\matA_{ij})} 
%= \alpha \cdot \frac{\abs{\matA_{ij}}}{\ONorm{\matA}} + (1-\alpha)\cdot \frac{\matA_{ij}^2}{\FNormS{\matA}},
$$  
where \quad $p_1 = \frac{\abs{\matA_{ij}}}{\ONorm{\matA}}$ \quad and \quad $p_2 = \frac{\matA_{ij}^2}{\FNormS{\matA}}$.
\end{theorem}
%
%====================================================
\begin{Proof}
Here we use the notations in Theorem \ref{thm:one_pass}.
Note that $t$-th elements of $S_1$ and $S_2$ are sampled independently with $\ell_1$ and $\ell_2$ probabilities, respectively.
We consider the following disjoint events:
\begin{eqnarray*}
\mathcal{E}_1: S_1(t) = (i,j,\matA_{ij}) \land S_2(t) \neq (i,j,\matA_{ij})\\
\mathcal{E}_2: S_1(t) \neq (i,j,\matA_{ij}) \land S_2(t) = (i,j,\matA_{ij})\\
\mathcal{E}_3: S_1(t) = (i,j,\matA_{ij}) \land S_2(t) = (i,j,\matA_{ij})\\
\mathcal{E}_4: S_1(t) \neq (i,j,\matA_{ij}) \land S_2(t) \neq (i,j,\matA_{ij})
\end{eqnarray*}
Let us denote the events $x_1: x \geq \alpha$ and $x_2: x < \alpha$. Clearly, $\prob{x_1} = \alpha, \prob{x_2} = 1-\alpha$.
Since the elements $S_1(t)$ and $S_2(t)$ are sampled independently, we have
\begin{eqnarray*}
\prob{\mathcal{E}_1} &=& \prob{S_1(t) = (i,j,\matA_{ij})}\prob{S_2(t) \neq (i,j,\matA_{ij})}= p_1(1-p_2)\\
\prob{\mathcal{E}_2} &=& (1-p_1)p_2\\
\prob{\mathcal{E}_3} &=& p_1 p_2\\
\prob{\mathcal{E}_4} &=& (1-p_1)(1-p_2)
\end{eqnarray*}
We note that $\alpha$ may be dependent on the elements of $S_3$ and $S_4$ (in Algorithm \ref{alg:alg_iterative_alpha}), but is independent of elements of $S_1$ and $S_2$. Therefore, events $x_1$ and $x_2$ are independent of the events $\mathcal{E}_j$, $j=1,2,3,4$. 
Thus, 
\begin{eqnarray*}
&&\prob{S(t) = (i,j,\matA_{ij})}\\
&=& \prob{(\mathcal{E}_1 \land x_1) \lor (\mathcal{E}_2 \land x_2) \lor \mathcal{E}_3}\\
&=& \prob{\mathcal{E}_1 \land x_1} + \prob{\mathcal{E}_2 \land x_2} + \prob{\mathcal{E}_3}\\
&=& \prob{\mathcal{E}_1}\prob{x_1} + \prob{\mathcal{E}_2}\prob{x_2} + \prob{\mathcal{E}_3}\\
&=& p_1(1-p_2)\alpha + (1-p_1)p_2(1-\alpha) + p_1p_2\\
&=& \alpha \cdot p_1 + (1-\alpha)\cdot p_2
\end{eqnarray*}
\end{Proof}  
%======================================================
%We note that the independent mulitsets $S_j$, $j=1,2,3,4$, in Algorithm 2 need not be generated using one pass over the data for Lemma \ref{thm:one_pass} to hold.
%

Note that, Theorem \ref{thm:one_pass} holds for any arbitrary $\alpha \in (0,1]$ in line 7 of Algorithm \ref{alg:alg_one_pass}, i.e., Algorithm \ref{alg:alg_iterative_alpha} is not essential for correctness of Theorem \ref{thm:one_pass}. We only need $\alpha$ to be independent of elements of $S_1$ and $S_2$. 
However, we use Algorithm \ref{alg:alg_iterative_alpha} to get an iterative estimate of $\alpha^*$ (Section \ref{proof:iterative_alpha}) in one pass over $\matA$. In this case, we need additional independent multisets $S_3$ and $S_4$ to `learn' the parameter $\alpha^*$. Algorithm \ref{alg:alg_one_pass} (without Algorithm \ref{alg:alg_iterative_alpha}) requires a memory twice as large required by $\ell_1$ or $\ell_2$ sampling. Using Algorithm \ref{alg:alg_iterative_alpha} this requirement is four times as large. However, in both the cases the asymptotic memory requirement remains the same $O(s)$.
%
%=============================================================
\subsection{Iterative Estimate of $\alpha^*$}\label{proof:iterative_alpha}
We obtain independent random multiset of triples $S_3$ and $S_4$, each containing $s$ elements from $\matA$ in one pass, in Algorithm \ref{alg:alg_one_pass}. We can create a sparse random matrix $\matX$, as shown in step 11 in Algorithm \ref{alg:alg_iterative_alpha}, that is an unbiased estimator of $\matA$. We use this $\matX$ as a proxy for $\matA$ to estimate the quantities we need in order to solve the optimization problem in (\ref{eqn:approx_alpha}). 
\begin{algorithm}[t]
\centerline{
\caption{Iterative estimate of $\alpha^*$}\label{alg:alg_iterative_alpha}
}
\begin{algorithmic}[1]
%--------------------------------------------------
\STATE \textbf{Input:} Multiset of triples $S_3$ and $S_4$ with $s$ elements each, number of iteration $\tau$, accuracy $\epsilon$,  $\FNormS{\matA}$, and $\ONorm{\matA}$. 
%--------------------------------------------------
%
\STATE Create empty multiset of triples $S$. % \leftarrow \emptyset$.
\STATE $\alpha_0 = 0.5$
\STATE \textbf{For} $k = 1\ldots \tau$
\STATE \quad $\matX \leftarrow \textbf{0}_{m\times n}$. 
\STATE \quad \textbf{For} $t = 1\ldots s$
\STATE \qquad Generate a uniform random number $x \in [0,1]$.\\
\STATE \qquad If $x \geq \alpha_{k-1}$, $S(t) \leftarrow S_3(t)$; else, $S(t) \leftarrow S_4(t)$.
\STATE \qquad $(i_t,j_t) \leftarrow S(t,1:2)$.
\STATE \qquad $p \leftarrow \alpha_{k-1}\cdot\frac{\abs{S(t,3)}}{\ONorm{\matA}} + (1-\alpha_{k-1})\cdot\frac{\abs{S(t,3)}^2}{\FNormS{\matA}}$ 
\STATE \qquad $\matX \leftarrow \matX + \frac{S(t,3)}{p\cdot s}e_{i_t}e_{j_t}^T$.
\STATE \quad \textbf{End}
%\STATE 
%\STATE \quad $\matA^{[k]} \leftarrow \frac{1}{s}\sum_{t=1}^s \frac{\matA_{i_t j_t}}{p_{i_t j_t}} \e_{i_t} \e_{j_t}^T$
\STATE \quad $\alpha_k \leftarrow$ $\tilde{\alpha}$ in (\ref{eqn:approx_alpha}) using $\matX$.
\STATE \textbf{End}
%
%\STATE \textbf{Output:} $\tilde{\alpha}$ in (\ref{eqn:approx_alpha}).
\STATE \textbf{Output:} $\alpha_{\tau}$.
%
%\STATE \textbf{Sampling operator:} $\mathcal{S}_{\Omega}: \mathbb{R}^{m \times n}\rightarrow \mathbb{R}^{m \times n}$ with
%
%$\mathcal{S}_{\Omega}\left(\matA\right) = \frac{1}{s}\sum_{t=1}^s \frac{\matA_{i_t j_t}}{p_{i_t j_t}} \e_{i_t} \e_{j_t}^T.$
%
\end{algorithmic}
\end{algorithm} 

%
%Therefore, using the above approximations for $\gamma(\alpha)$ and $\rho^2(\alpha)$ we can determine the value of $\alpha$ %for which $s(\alpha)$ achieves its minimum, i.e., for given $\epsilon$ and $\delta$,
%
\begin{eqnarray}\label{eqn:approx_alpha}
\tilde{\alpha}:  \min_{\alpha \in (0,1]}\left\{ \left( \tilde{\rho}^2(\alpha) + \tilde{\gamma}(\alpha)\epsilon \TNorm{\matX}/3\right) \right\} 
\end{eqnarray}
where, for all $(i,j) \in S(:,1:2)$ 
$$
\tilde{\xi}_{ij} = {\FNormS{\matX}}/\left (\frac{\alpha \cdot  \FNormS{\matX}}{\abs{\matX_{ij}}\cdot \ONorm{\matX}}+ (1-\alpha)\right ),$$ %for $(i,j) \in S(:,1:2),$
$$
\tilde{\rho}^2(\alpha) = \max\left\{\max_i \sum_{j=1}^{n}\tilde{\xi}_{ij},\max_j \sum_{i=1}^{m}\tilde{\xi}_{ij}\right\},
%\text{ for } (i,j) \in S(:,1:2)
$$
$$
\tilde{\gamma}(\alpha) = \max_{ij}\left\{\frac{\ONorm{\matX}}{\alpha + (1-\alpha)\frac{\ONorm{\matX}\cdot \abs{\matX_{ij}}}{\FNormS{\matX}}}\right\} + \FNorm{\matX}. 
%\text{ for } (i,j) \in S(:,1:2).
$$
We note that $\ZNorm{\matX} \leq s$. We can compute the quantities $\tilde{\rho}(\alpha)$ and $\tilde{\gamma}(\alpha)$, for a fixed $\alpha$, using $O(s)$ memory. We consider $\varepsilon = \epsilon \cdot \TNorm{\matX}$ to be the given accuracy. 
%Thus, in one pass over $\matA$ and using $O(\max\{s, m+n\})$ memory we can find $\tilde{\alpha}$ in eqn (\ref{eqn:approx_alpha}).
%=============================================================

\section{Fast Approximation of PCA}
\label{sec:PCA}
Here, we discuss a provable algorithm (Algorithm \ref{alg:alg_approx_PCA}) to speed up computation of PCA applying element-wise sampling. We sparsify a given centered data $\matA$ to produce a sparse unbiased estimator $\tilde{\matA}$ by sampling $s$ elements in i.i.d. trials according to our hybrid-$(\ell_1,\ell_2)$ distribution in (\ref{eqn:probability}). Computation of rank-truncated SVD on sparse data is fast, and we consider the right singular vectors of $\tilde{\matA}$ as the approximate principal components of $\matA$. 
Naturally, more samples produce better approximation. However, this reduces sparsity, and consequently we lose the speed advantage. 
%$\matP_A = \matA\tilde{\matV}_k\tilde{\matV}_k^T$ is the projected data onto the space spanned by approximate PCs. In this case, we assume to have access to all the data elements of $\matA$, and we can compute $\alpha^*$ and sample elements according to our optimal hybrid-$(\ell_1,\ell_2)$ distribution to produce $\tilde{\matA}$.
% 
\begin{algorithm}[t]
\centerline{
\caption{Fast Approximation of PCA}\label{alg:alg_approx_PCA}
}
\begin{algorithmic}[1]
%--------------------------------------------------
\STATE \textbf{Input:} Centered data $\matA  \in \mathbb{R}^{m \times n}$, sparsity parameter $s>0$, and rank parameter $k$.
%--------------------------------------------------
%\STATE Sparsify $\matA$ to produce an unbiased estimator $\tilde{\matA}$ by sampling (and rescaling) $s$ elements in i.i.d. trials according to distribution (\ref{eqn:probability}).
%
\STATE Produce sparse unbiased estimator $\tilde{\matA}$ from $\matA$, in $s$ i.i.d. trials using Algorithm \ref{alg:alg1}. 
\STATE Perform rank truncated SVD on sparse matrix $\tilde{\matA}$, i.e., $[\tilde{\matU}_k, \tilde{\matD}_k, \tilde{\matV}_k]$ = SVD($\tilde{\matA}$, $k$). 
%
%\STATE $\matP_A = (\matA\tilde{\matV}_k)(\matA\tilde{\matV}_k)^{\dagger}\matA$, where 
%$\matX^\dagger$ is the pseudo-inverse of $\matX$.
%\STATE $\tilde{\matP}_A = (\tilde{\matA}\tilde{\matV}_k)(\tilde{\matA}\tilde{\matV}_k)^{\dagger}\tilde{\matA}$.
%
%\STATE \textbf{Output:}  ${\matP}_A$ and $\tilde{\matP}_A$.
\STATE \textbf{Output:}  $\tilde{\matV}_k$ (columns of $\tilde{\matV}_k$ are the ordered PCA's).
%
%$\tilde{\matP}_A = \tilde{\matU}_k(\tilde{\matU}_k^T\tilde{\matA})$
%
%$\matP_A = \tilde{\matU}_k(\tilde{\matU}_k^T\matA$), and $\tilde{\matP}_A = \tilde{\matU}_k(\tilde{\matU}_k^T\tilde{\matA})$
%
\end{algorithmic}
\end{algorithm} 

%
%We can use Algorithm 4 also in a situation where we can have access to only $s$ elements of $\matA$. We apply Algorithm 2 along with Algorithm 3 to produce sparse unbiased estimator $\tilde{\matA}$ using estimated optimal hybrid sampling. We then apply rank-truncated SVD (step 3 of Algorithm 4) on $\tilde{\matA}$ to approximate PCs. 
%$\tilde{\matP}_A$ gives the projection of the unbiased estimator onto the space spanned by the approximate PCs.\\
%
%Note that,
%(WRONG !!) $\matP_A = (\matA\tilde{\matV}_k)(\matA\tilde{\matV}_k)^{\dagger}\matA = \matA\tilde{\matV}_k\tilde{\matV}_k^T$, and 
%$\tilde{\matP}_A = (\tilde{\matA}\tilde{\matV}_k)(\tilde{\matA}\tilde{\matV}_k)^{\dagger}\tilde{\matA} = \tilde{\matA}_k$.
%
Theorem \ref{thm:approx_PCA} shows the quality of approximation of principal components produced by Algorithm \ref{alg:alg_approx_PCA}. 
%
%=================================
\begin{theorem}\label{thm:approx_PCA}
Let $\matA \in \mathbb{R}^{m \times n}$ be a given matrix, and $\tilde{\matA}$ be a sparse sketch produced by Algorithm \ref{alg:alg1}. Let $\tilde{\matV}_k$ be the PCA's of $\tilde{\matA}$ computed in step 3 of Algorithm \ref{alg:alg_approx_PCA}. Then
\begin{eqnarray*}
\nonumber \FNormS{\matA - \matA\tilde{\matV}_k\tilde{\matV}_k^T} &\leq& \FNormS{\matA - \matA_k} + \frac{4\FNormS{\matA_k}}{\sigma_k(\matA)}\TNorm{\matA - \tilde{\matA}}\\
%
%\end{eqnarray*}
%\begin{eqnarray*}
%
\FNorm{\matA_k - \tilde{\matA}_k} &\leq& \sqrt{8k}\cdot \left(\TNorm{\matA - {\matA}_k} + \TNorm{\matA - \tilde{\matA}}\right)\\
%
%\end{eqnarray*}
%\begin{eqnarray*}
%
\FNorm{\matA - \tilde{\matA}_k} &\leq& \FNorm{\matA - {\matA}_k} 
+ \sqrt{8k}\cdot \left(\TNorm{\matA - {\matA}_k} + \TNorm{\matA - \tilde{\matA}}\right)
\end{eqnarray*}
%
%where, $\sigma_k(\matA)$ is $k$-th largest singular value of $\matA$.
\end{theorem}
The first inequality of Theorem \ref{thm:approx_PCA} bounds the approximation of projected data onto the space spanned by top $k$ approximate PCA's. The second and third inequalities measure the quality of $\tilde{\matA}_k$ as a surrogate for $\matA_k$ and the quality of projection of sparsified data onto approximate PCA's, respectively.

Proofs of first two inequalities of Theorem \ref{thm:approx_PCA} follow from Theorem 5 and Theorem 8 of \cite{AM01}, respectively. The last inequality follows from the triangle inequality. The last two inequalities above are particularly useful in cases where $\matA$ is inherently low-rank and we choose an appropriate $k$ for approximation, for which $\TsNorm{\matA - \matA_k}$ is small.

\section{Experiments}\label{experiments}
In this section we perform various element-wise sampling experiments on synthetic and real data to show how well the sparse sketches preserve the structure of the original data, in spectral norm. Also, we show results on the quality of the PCA's derived from sparse sketches. 
%First we briefly describe the synthetic and real datasets we used here.
%We measure some of the matrix metrics that were defined in~\cite{AKL13} for various data, and list them in Table \ref{fig:table_data_summary} in Appendix.
%Various characteristics of the datasets are summarized in Table \ref{fig:table_data_summary}. 

% ===========================
\subsection{Algorithms for Sparse Sketches}

We use Algorithm \ref{alg:alg1} as a prototypical algorithm to produce sparse sketches from a given matrix via various sampling methods. Note that, we can plug-in any element-wise probability distribution in Algorithm \ref{alg:alg1} to produce (unbiased) sparse matrices. We construct sparse sketches via our optimal hybrid-($\ell_1,\ell_2$) sampling, along with other sampling methods related to extreme choices of $\alpha$, such as, $\ell_1$ sampling for $\alpha=1$. 
Also, we use \textit{element-wise leverage scores} (\cite{BCSW14}) for sparsification of \textit{low-rank} data. Element-wise leverage scores are used in the context of \textit{low-rank matrix completion} by \cite{BCSW14}. Let $\matA$ be a $m \times n$ matrix of rank $\rho$, and its SVD if given by $\matA = \matU\matSig\matV^T$. Then, we define $\mu_i$ (row leverage scores), $\nu_j$ (column leverage scores), and element-wise leverage scores $p_{lev}$ as follows:
\begin{eqnarray*}
\mu_i = \TNormS{\matU_{(i)}}, \quad \nu_j = \TNormS{\matV_{(j)}}, \quad 
p_{lev} = \frac{\mu_i + \nu_j}{(m+n)\rho}, \quad i\in [m], j \in [n]
\end{eqnarray*}
%
%for $i=1,...,m$, $j=1,..., n$.\\
Note that $p_{lev}$ is a probability distribution on the elements of $\matA$. Leverage scores become uniform if the matrix $\matA$ is full rank. %But our hybrid-$(\ell_1,\ell_2)$ sampling is applicable for general matrices. 
We use $p_{lev}$ in Algorithm \ref{alg:alg1} to produce sparse sketch $\tilde{\matA}$ of a low-rank data $\matA$.

\subsubsection{Experimental Design for Sparse Sketches}

We compute the theoretical optimal mixing parameter $\alpha^*$ by solving eqn (\ref{eqn:opt_alpha}) for various datasets. We compare this $\alpha^*$ with the theoretical condition derived by \cite{AKL13} (for cases when $\ell_1$ sampling outperforms $\ell_2$ sampling). 
We verify the accuracy of $\alpha^*$ by measuring the quality of the sparse sketches $\tilde{\matA}$, $\mathcal{E} = \TsNorm{\matA - \tilde{\matA}}/\TsNorm{\matA}$ for various sampling distributions. Let $\mathcal{E}_h$, $\mathcal{E}_1$, and $\mathcal{E}_{lev}$ denote the quality of sparse sketches produced via optimal hybrid sampling, $\ell_1$ sampling, and element-wise leverage scores $p_{lev}$, respectively. We compare $\mathcal{E}_h$, $\mathcal{E}_1$, and $\mathcal{E}_{lev}$ for various sample sizes for real and synthetic datasets.  

% ===========================
\subsection{Algorithms for Fast PCA}

We compare three algorithms for computing PCA of the centered data. Let the actual PCA of the original data be $\mathcal{A}$. We use Algorithm \ref{alg:alg_approx_PCA} to compute approximate PCA via our optimal hybrid-$(\ell_1,\ell_2)$ sampling. Let us denote this approximate PCA by $\mathcal{H}$. Also, we compute PCA of a Gaussian random projection of the original data to compare the quality of $\mathcal{H}$. 
Let $\matA_G = \matG \matA \in \mathbb{R}^{r \times n}$, where $\matA \in \mathbb{R}^{m \times n}$ is the original data, and $\matG$ is a $r \times m$ standard Gaussian matrix. Let the PCA of this random projection $\matA_G$ be $\mathcal{G}$. Also, let $T_a$, $T_h$, and $T_G$ be the computation time (in milliseconds) for $\mathcal{A}$, $\mathcal{H}$, and $\mathcal{G}$, respectively.

\subsubsection{Experimental Design for Fast PCA}

We compare the visual quality of $\mathcal{A}$, $\mathcal{H}$, and $\mathcal{G}$ for image datasets. Also, we compare the computation time $T_a$, $T_h$, and $T_G$ for these datasets.

% ===========================
\subsection{Description of Data}

In this section we describe the synthetic and real datasets we use in our experiments.

\subsubsection{Synthetic Data}
%\textbf{Synthetic Data:}
%
We construct a binary $500 \times 500$ image data $\matD$ (see Figure \ref{fig:1}). We add random noise to perturb the elements of the `pure' data $\matD$. Specifically, we construct a $500 \times 500$ noise matrix $\textbf{N}$  whose elements $\textbf{N}_{ij}$ are drawn i.i.d from  Gaussian with mean zero and standard deviation $\sigma$. We use two different values for $\sigma$ in our experiments: $\sigma=0.05$ and $\sigma=0.10$. For each $\sigma$, we note the following ratios:
$$\text{Noise-to-signal energy ratio} = {\FNorm{\textbf{N}}}/{\FNorm{\matD}},$$
$$\text{ Spectral ratio} = {\TNorm{\textbf{N}}}/{\sigma_{k}(\matD)},$$   
where $\sigma_k(\matD)$ is the $k$-th largest singular value of $\matD$. 
%Spectral ratio is an indicator of how much the noise may distort the $k$-th strongest linear trend in the data. 
%
For $\sigma=0.05$ and $\sigma=0.10$, average  Noise-to-signal energy ratio are $0.44$ and $0.88$, average Spectral ratio are $0.09$  and $0.17$, and average maximum absolute values of noise turn out to be $0.25$ and $0.50$, respectively. 
%We set $k=5$ (rank of $\matD$) for synthetic data $\matA = \matD + \textbf{N}$. 
We denote noisy data by $\matA_{0.05}$ (respectively $\matA_{0.1}$) when $\matD$ is perturbed by $\textbf{N}$ whose elements $\textbf{N}_{ij}$ are drawn i.i.d from a Gaussian distribution with mean zero and $\sigma = 0.05$ (respectively $\sigma = 0.1$).
%
%Table \ref{fig:table1} lists these quantities. 

\subsubsection{TechTC Datasets}
%\textbf{TechTC Datasets}(\cite{GM04}):
%
These datasets (\cite{GM04}) are bag-of-words features for document-term data describing two topics (ids). We choose four such datasets: TechTC1 with ids 10567 and 11346, TechTC2 with ids 10567 and 12121, TechTC3 with ids 11498 and 14517, TechTC4 with ids 11346 and 22294. Rows represent documents and columns are the words. We preprocessed the data by removing all the words of length four or smaller, and then normalized the rows by dividing each row by its Frobenius norm. 
%All these datasets show reasonable separation between two classes of topics when projected onto the space spanned by top two principal components. 
The following table lists the dimension of the TechTC datasets.
%  ===============================================
\begin{table}[!h]
\begin{center}
    \begin{tabular}{| c | c | c | c |}
   \hline
 Dimension ($m \times n)$ &  $m$ &  $n$   \\ \hline \hline  
 TechTC1 &  139  &  15170  \\ \hline
 TechTC2 &   138  &   11859  \\ \hline
 TechTC3 &   125  &  15485  \\ \hline
 TechTC4 &   125  &  14392  \\ \hline
   \end{tabular}
%\caption{Values of $\tilde{\alpha}$ (estimated $\alpha^*$ using Algorithm 3) for various data sets using one pass over the elements of data and $O(s)$ memory.}
\caption{Dimension of TechTC datasets}
%\label{fig:alpha_tilde}
\end{center}
\end{table}
%  ===============================================
%This is to reduce the bias towards larger documents.\\
%

\subsubsection{Handwritten Digit Data}
%\textbf{Digit Data}(\cite{Hull94}):
%
A dataset (\cite{Hull94}) of three handwritten digits: six (664 samples), nine (644 samples), and one (1005 samples). Pixels are treated as features, and pixel values are normalized in [-1,1]. Each $16 \times 16$ digit image is first represented by a column vector by appending the pixels column-wise. Then, we use the transpose of this column vector to form a row in the data matrix. The number of rows $m = 2313$, and columns $n=256$.
%This dataset shows good separation of three classes of digits when projected onto the space spanned by top three principal components.

\subsubsection{Stock Data}
%\textbf{Stock Data:}
%
We use a stock market dataset (S\&P) containing prices of 1218 stocks collected between 1983 and 2011. This temporal dataset has 7056 snapshots of stock prices. Thus, we have $m = 1218$ and $n = 7056$. \\

%=====================
%\begin{figure}[!h]
%\centering
%	\includegraphics[height=3.0cm,width=4.5cm]{Synthetic_data_im.eps}
%	\label{fig:data_im}
%	\qquad  
%	\includegraphics[height=3.5cm,width=5.8cm]{Synthetic_k_5_data_mesh_noisy10.eps}
%\caption{(left) Synthetic noiseless binary data $\matD$; (right) mesh view of noisy data $\matA0.1$.} 
%\label{fig:1}
%\end{figure}
%=====================

%Dimensions of datasets used here are $m \times n$. 

We provide summary statistics for all the datasets in Table \ref{fig:table_data_summary}. 
In order to compare our results with \cite{AKL13} we review the matrix metrics that they use. 
Let the numeric density of matrix $\matX$ be 
$\text{nd}(\matX) = \ONormS{\matX}/\FNormS{\matX}.$
Clearly, $\text{nd($\matX)$}\leq \ZNorm{\matX}$, with equality holding for zero-one matrices. The row density skew of $\matX$ is defined as 
$$\text{rs}_0(\matX) = \frac{\max_i\ZNorm{\matX_{(i)}}}{\ZNorm{\matX}/m},$$ i.e., the ratio between number of non-zeros in the densest row and the average number of non-zeros per row. The numeric row density skew, 
$$\text{rs}_1(\matX) = \frac{\max_i\ONorm{\matX_{(i)}}}{\ONorm{\matX}/m},$$ is a smooth analog of $\text{rs}_0(\matX)$.
\cite{AKL13} assumed that $m \leq n$ without loss of generality, and for simplicity, $\max_i \XsNorm{\matX_{(i)}} \geq \max_i \XsNorm{\matX^{(i)}},$ for all $\xi \in \{0,1,2\}$.
We notice that, although the Digit dataset does not satisfy the above conditions, its transpose does. We can work on the transposed dataset without loss of generality, and hence we take note of $\text{rs}_0$ and $\text{rs}_1$ of the transposed Digit data.

%=====================
\begin{table}[!h]
\begin{center}
    \begin{tabular}{|  l | l | l | l | l |  l |}
   \hline
  &  $\ZNorm{\matX}$ & \quad nd & $\text{rs}_0$ & $\text{rs}_1$ \\ \hline \hline  
$\matA_{0.05}$ & 2.5e+5 & 4.4e+4 & 1 & 2.66 \\ \hline
$\matA_{0.10}$ & 2.5e+5 & 9.2e+4 & 1 & 1.95 \\ \hline
TechTC1 & 37831 & 12204 & 5.14 & 2.18 \\ \hline
TechTC2 & 29334 & 9299 & 3.60 & 2.10 \\ \hline
TechTC3 & 47304 & 14201 & 7.23 & 2.31 \\ \hline
TechTC4 & 35018 & 10252 & 4.99 & 2.25 \\ \hline
Digit & 5.9e+5 & 5.1e+5 & 1 & 1.3  \\ \hline
Stock & 5.5e+6 & 6.5e+3 & 1.56 & 1.1e+03  \\ \hline
    \end{tabular}
\caption{Summary statistics for the data sets}
\label{fig:table_data_summary}
\end{center}
\end{table}
%=====================

\subsection{Results}
We report all the results based on an average of five independent trials. We observe a small variance of the results. 
\subsubsection{Quality of Sparse Sketch}

We first note that three sampling methods $\ell_1$, $\ell_2$, and hybrid-$(\ell_1,\ell_2)$, perform identically on noiseless data $\matD$. 
%Statistics of the noisy data $\matA$ in Table \ref{fig:table_data_summary}, along with condition (\ref{eqn:L1>L2}), suggest that %$L_1$ sampling $(\alpha=1)$ is likely to  perform much better than $L_2$ on $\matA$. 
%
We report the total probability of sampling noisy elements in $\matA = \matD + \textbf{N}$ (elements which are zeros in $\matD$). $\ell_1$ sampling shows the highest susceptibility to noise, whereas, small-valued noisy elements are suppressed in $\ell_2$. Hybrid-$(\ell_1,\ell_2)$ sampling, with $\alpha < 1$, samples mostly from true data elements, and thus captures the low-rank structure of the data better than $\ell_1$. The optimal mixing parameter $\alpha^*$ maintains the right balance between $\ell_2$ sampling and $\ell_1$ regularization and gives the smallest sample size to achieve a desired accuracy.
%Also, $\alpha \neq 0$ dependence on $L_1$ ensures $L_1+L_2$ is never worse than a factor $1/\alpha$ even in case of picking an %extremely small element. 
%
Table \ref{fig:alpha_opt} summarizes $\alpha^*$ for various data sets. 
\iffalse % ====== COMMENTED OUT =========
In order to compare our results with \cite{AKL13} we review  the matrix metrics that they use. 
Let the numeric density of matrix $\matX$ be 
$\text{nd}(\matX) = \ONormS{\matX}/\FNormS{\matX}.$
Clearly, $\text{nd($\matX)$}\leq \ZNorm{\matX}$, with equality holding for zero-one matrices. The row density skew of $\matX$ is defined as 
$$\text{rs}_0(\matX) = \frac{\max_i\ZNorm{\matX_{(i)}}}{\ZNorm{\matX}/m},$$ i.e., the ratio between number of non-zeros in the densest row and the average number of non-zeros per row. The numeric row density skew, 
$$\text{rs}_1(\matX) = \frac{\max_i\ONorm{\matX_{(i)}}}{\ONorm{\matX}/m},$$ is a smooth analog of $\text{rs}_0(\matX)$.
%
\cite{AKL13} assumed that $m \leq n$ without loss of generality, and for simplicity, $\max_i \XsNorm{\matX_{(i)}} \geq \max_i \XsNorm{\matX^{(i)}},$ for all $\xi \in \{0,1,2\}$.
%
We notice that, although the Digit dataset does not satisfy the above conditions, its transpose does. We can work on the transposed dataset without loss of generality, and hence we take note of $\text{rs}_0$ and $\text{rs}_1$ of the transposed Digit data.
\fi % ====== END OF COMMENTED OUT =========
\cite{AKL13} argued that, as long as $\text{rs}_0(\matX)\geq \text{rs}_1(\matX)$, $\ell_1$ sampling is better than $\ell_2$ (even with truncation). Our results on $\alpha^*$ in Table \ref{fig:alpha_opt} confirm this condition. Moreover, our method can derive the right blend of $\ell_1$ and $\ell_2$ sampling even when the above condition fails. In this sense, we generalize the results of \cite{AKL13}.
%  ===============================================
\begin{table}[!h]
\begin{center}
    \begin{tabular}{|  c | c | c | c |}
   \hline
 \quad  & $\epsilon=0.05$ & $\epsilon=0.75$ & $\text{rs}_0 \geq \text{rs}_1$  \\ \hline \hline  
%$\alpha^*$,  & 0.62 & 0.63 & \qquad 1 & \qquad 1 & \qquad 1 & \qquad 1 & 0.20 \\\hline 
%$\alpha^*$, $\epsilon=0.75$ & 0.69 & 0.70 & \qquad 1 & \qquad 1 & \qquad 1 & \qquad 1 & 0.74 \\\hline 
$\matA_{0.05}$ &  0.62 &  0.69 &  no \\ \hline 
$\matA_{0.1}$ &  0.63 &  0.70 &   no \\ \hline
TechTC1 &  1 &  1 &   yes \\ \hline
TechTC2 &  1 &  1 &   yes  \\ \hline
TechTC3 &  1 &  1 &   yes  \\ \hline
TechTC4 &  1 &  1 &   yes  \\ \hline
Digit &  0.20 &  0.74 &   no  \\ \hline
Stock &  0.74 &  0.75 &  no   \\ \hline
   \end{tabular}
\caption
{
$\alpha^*$ for various data sets ($\epsilon$ is the desired relative-error accuracy). The last column compares $\alpha^*$ with the condition established by \cite{AKL13}. Whenever $\text{rs}_0 \geq \text{rs}_1$, \cite{AKL13} show that $\ell_1$ sampling is always better than $\ell_2$ sampling, and we find $\alpha^*=1$ ($\ell_1$ sampling). However, when $\text{rs}_0 < \text{rs}_1$, $\alpha^*<1$ and our hybrid sampling is strictly better.
}
\label{fig:alpha_opt}
\end{center}
\end{table}
%  ===============================================

\begin{figure}[!h]
\centering
\begin{subfigure}[b]{0.4\textwidth}
	\includegraphics[height=4.0cm,width=6.5cm]{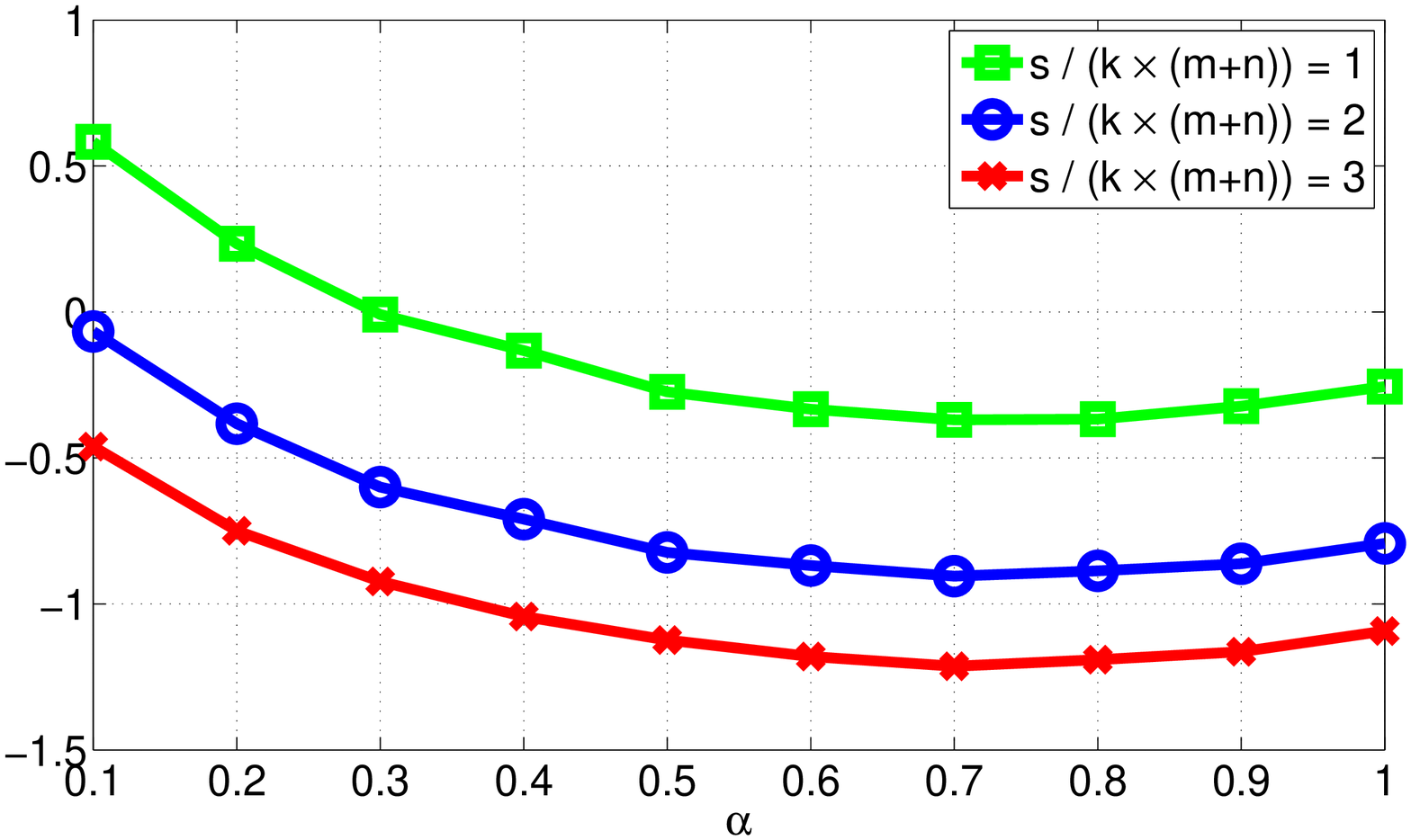}
\caption{$\matA_{0.05}$}
\end{subfigure}
\begin{subfigure}[b]{0.4\textwidth}
	\includegraphics[height=4.0cm,width=6.5cm]{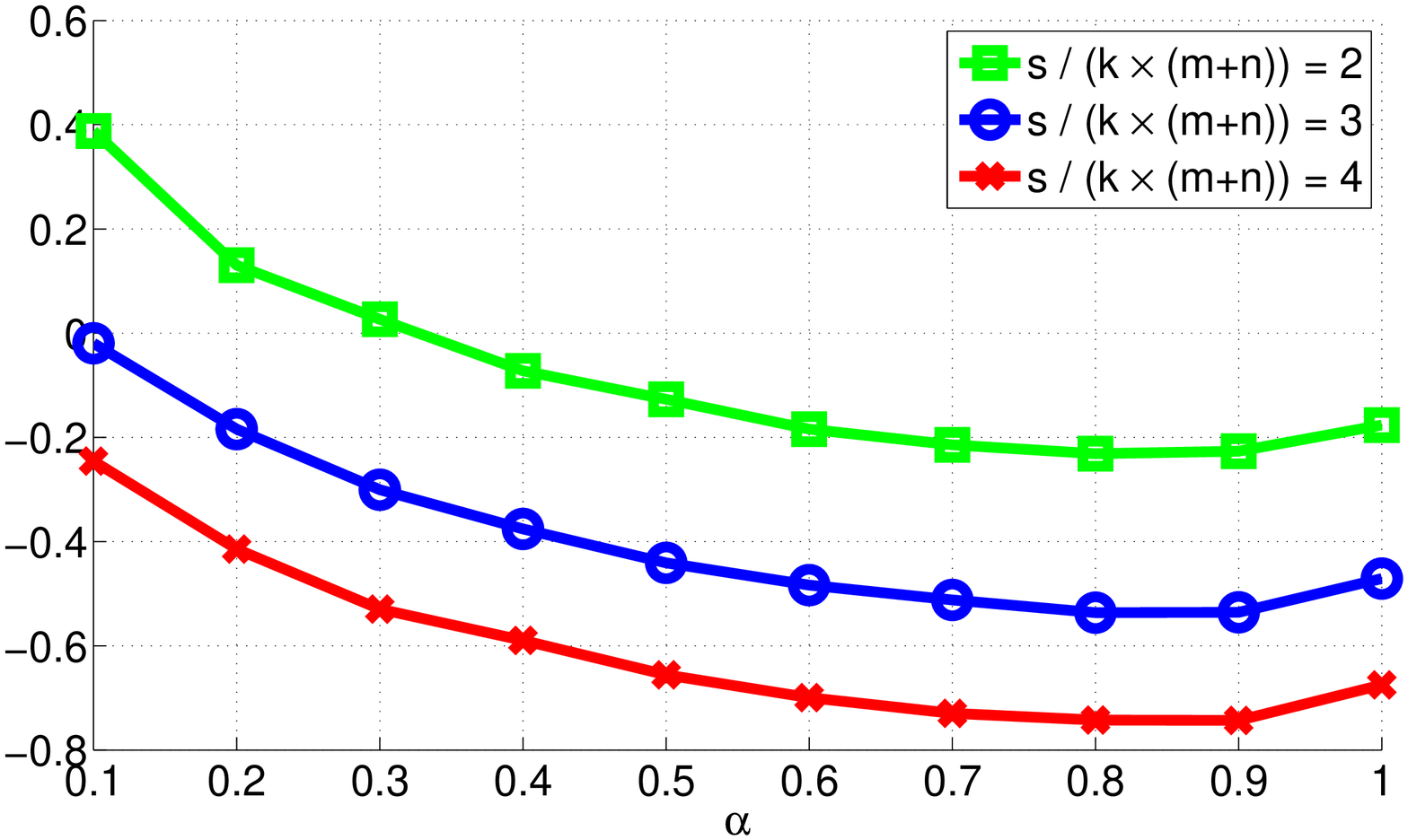}
\caption{$\matA_{0.1}$}
\end{subfigure}
%\begin{subfigure}[b]{0.4\textwidth}
%
%	\includegraphics[height=3.4cm,width=5.6cm]{TECHTC_Exp_10567_11346_compress_k12_plot.eps}
%
%	\includegraphics[height=3.4cm,width=5.6cm]{TECHTC_Exp_10567_12121_compress_k12_plot.eps}
%
%	\includegraphics[height=3.4cm,width=5.6cm]{TECHTC_Exp_11498_14517_compress_k9_plot.eps}
%
%	\includegraphics[height=3.4cm,width=5.6cm]{TECHTC_Exp_11346_22294_compress_k13_plot.eps}
%\end{subfigure}
\begin{subfigure}[b]{0.4\textwidth}
	\includegraphics[height=4.0cm,width=6.5cm]{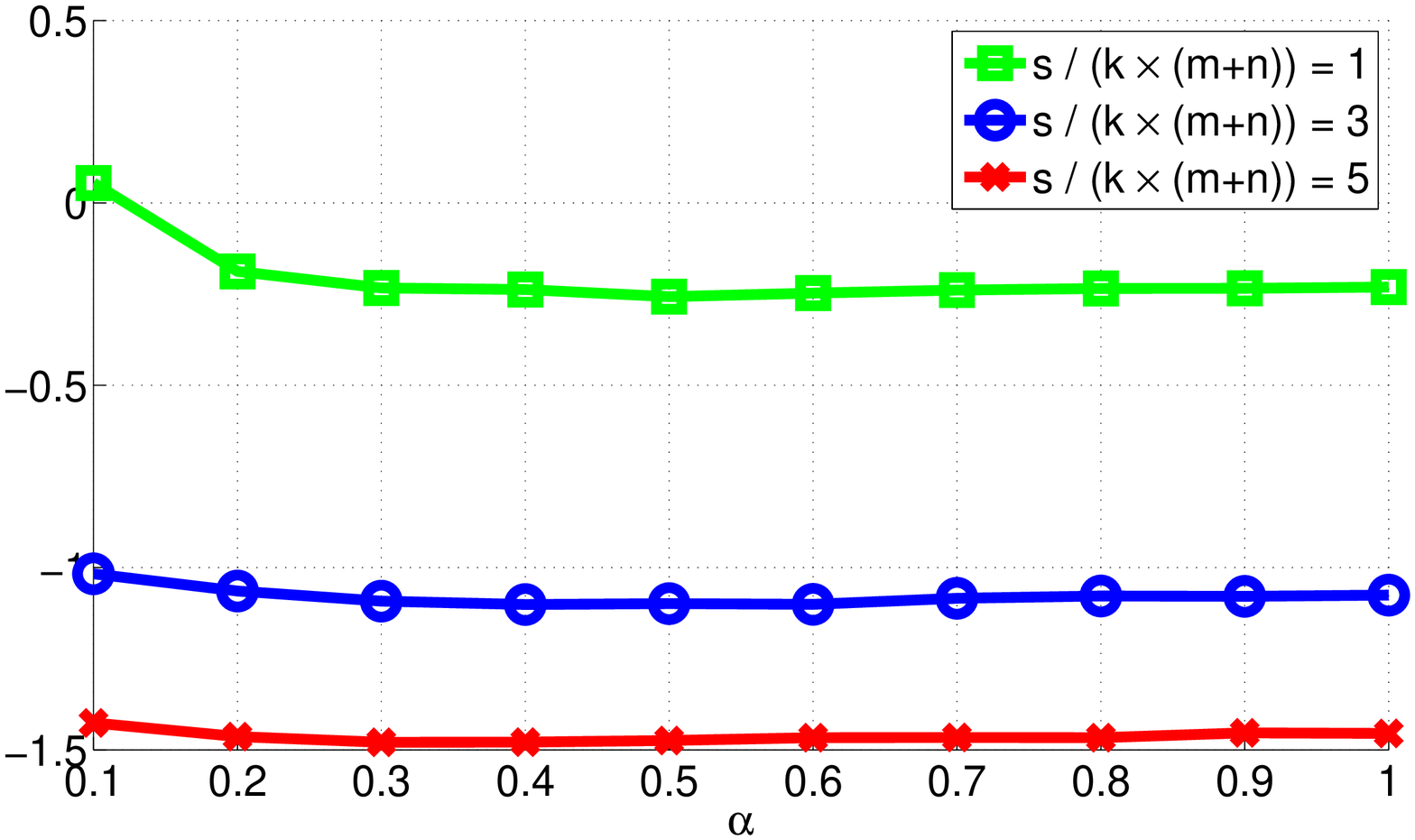}
\caption{Digit}
\end{subfigure}
\begin{subfigure}[b]{0.4\textwidth}
	\includegraphics[height=4.0cm,width=6.5cm]{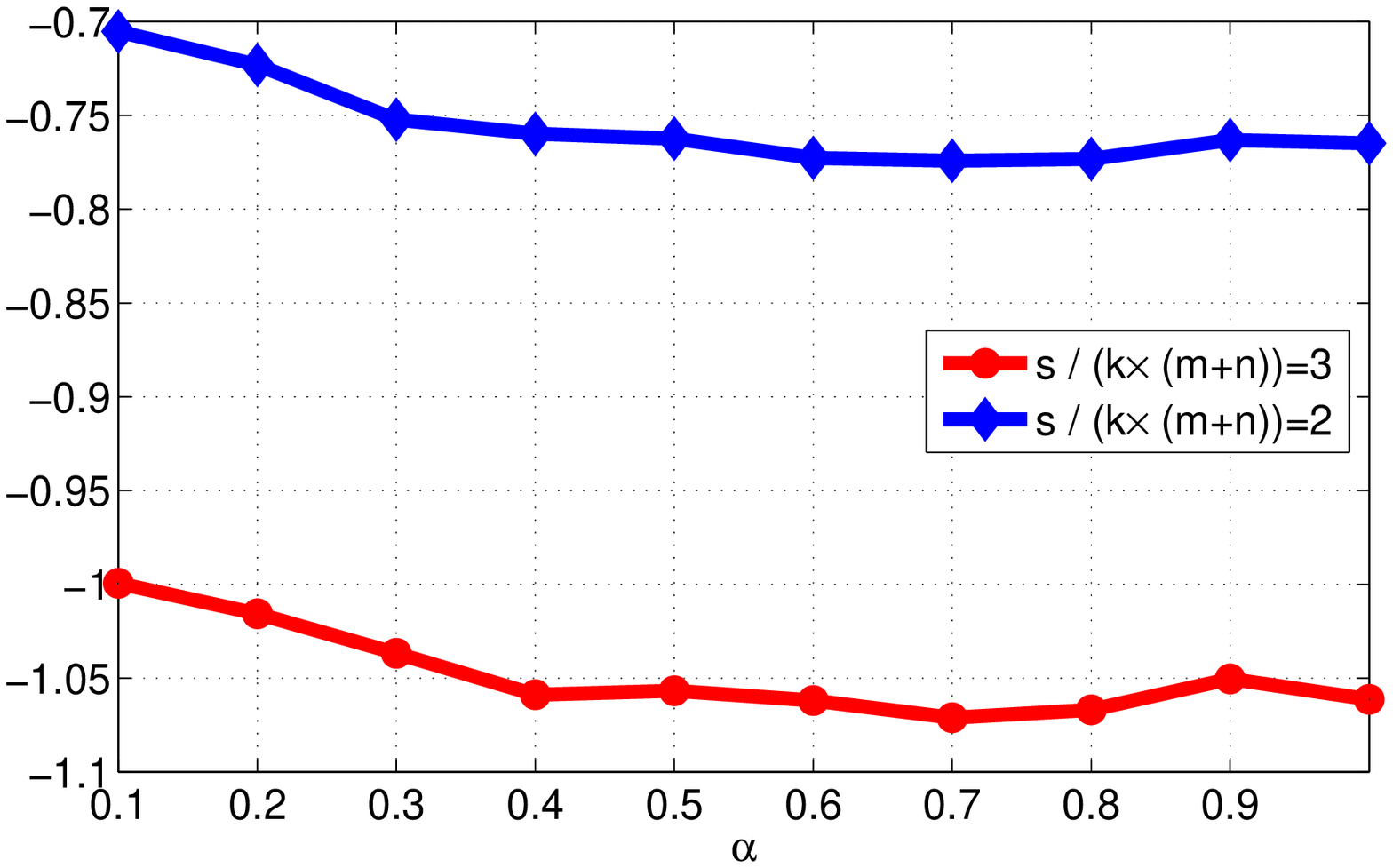}
\caption{Stock}
\end{subfigure}

\caption{Approximation quality of sparse sketch $\tilde{\matA}$: hybrid-$(\ell_1,\ell_2)$ sampling, for various $\alpha$ and different sample size $s$, are shown. $x$-axis is $\alpha$, and $y$-axis plots ${\TsNorm{\matA - \tilde{\matA}}}/{\TsNorm{\matA}}$ (in log$_2$ scale such that larger negative values indicate better quality). 
Each figure corresponds to a dataset: (a) $\matA_{0.05}$, (b) $\matA_{0.1}$, (c) Digit, and (d) Stock.
We set $k=5$ for synthetic data, $k=3$ for Digit data, and $k=1$ for Stock data. Choice of $k$ is close to the stable rank of the data.
%$y$-values are in log$_2$ scale such that larger negative values indicate better quality.
}
\label{fig:compress_approx}
\end{figure}

Figure \ref{fig:compress_approx} plots  $\mathcal{E} = {\TsNorm{\matA - \tilde{\matA}}}/{\TsNorm{\matA}}$ for various values of $\alpha$ and sample size $s$ for various datasets. It clearly shows our optimal hybrid sampling is superior to $\ell_1$  or $\ell_2$ sampling.

% ============================================
%
% ============================================

%  ===============================================
%We also perform experiments on real and synthetic data in order to compare the quality of approximation of $\tilde{\matA}$ using 
We also compare the quality of sparse sketches produced via our hybrid sampling with that of $\ell_2$ sampling with truncation. We use two predetermined truncation parameters, $\epsilon = 0.1$ and $\epsilon = 0.01$, for $\ell_2$ sampling.
First, $\ell_2$ sampling without truncation turns out to be the worst for all datasets. $\ell_2$ with $\epsilon=0.01$ appears to produce sparse sketch $\tilde{\matA}$ that is as bad as $\ell_2$ without truncation for $\matA_{0.1}$ and $\matA_{0.05}$. However, $\ell_2$ with $\epsilon = 0.1$ shows better  performance than hybrid sampling, for $\matA_{0.1}$ and $\matA_{0.05}$, because this choice of $\epsilon$ turns out to be an appropriate threshold to zero-out most of the noisy elements. 
We must point out that, in this example, we control the noise, and we know what a good threshold may look like. However, in reality we have no control over the noise. Therefore, choosing the right threshold for $\ell_2$, without any prior knowledge, is an improbable task. For real datasets, it turns out that hybrid-$(\ell_1,\ell_2)$-hybrid sampling using $\alpha^*$ outperforms $\ell_2$ sampling with the predefined thresholds for various sample sizes.

We compare the quality of Algorithm \ref{alg:alg_iterative_alpha} producing an iterative estimate of $\alpha^*$ in a very restricted set up, i.e., one pass over the elements of data using $O(s)$ memory. Table \ref{fig:alpha_tilde} lists $\tilde{\alpha}$, the estimated $\alpha^*$, for some of the datasets, for two choices of $s$ using 10 iterations. We compare these values with the plots in Figure \ref{fig:compress_approx} where the results are generated without any restriction of size of memory or number of pass over the elements of the datasets. 
\begin{table}[!h]
\begin{center}
    \begin{tabular}{| c | c | c |}
   \hline
  & $\frac{s}{k\cdot (m+n)} = 2$ & $\frac{s}{k\cdot (m+n)} = 3$   \\ \hline \hline  
$\matA_{0.05}, k=5$ &  0.54 &  0.48 \\ \hline
$\matA_{0.1}, k=5$ &  0.55 &  0.5 \\ \hline
%TechTC1, $k=2$ &  \qquad 0.93 & \qquad  1 \\ \hline
%TechTC2, $k=2$ &  \qquad 1 & \qquad 1 \\ \hline
%TechTC3, $k=2$ &  \qquad 0.97 & \qquad 0.85 \\ \hline
%TechTC4, $k=2$ &  \qquad 0.79 & \qquad 1 \\ \hline
Digit, $k=3$ &  0.69 &  0.89 \\ \hline
Stock, $k=1$ &  1 &  1 \\ \hline
   \end{tabular}
\caption{Values of $\tilde{\alpha}$ (estimated $\alpha^*$ using Algorithm \ref{alg:alg_iterative_alpha}) for various data sets using one pass over the elements of data and $O(s)$ memory. We use $\epsilon=0.05$, $\delta=0.1$.}
\label{fig:alpha_tilde}
\end{center}
\end{table}
%  ===============================================

Finally, we compare our hybrid-$(\ell_1,\ell_2)$ sampling with \textit{element-wise leverage score} sampling (similar to \cite{BCSW14}) to produce quality sparse sketches from low-rank matrices. For this, 
we construct a $500 \times 500$ low-rank \textit{power-law} matrix, similar to \cite{BCSW14}, as follows:
$\matA_{pow} = \matD\matX\matY^T\matD$, where, matrices $\matX$ and $\matY$ are $500 \times 5$ i.i.d. Gaussian $\mathcal{N}(0,1)$ and  $\matD$ is a diagonal matrix with power-law decay, $\matD_{ii} = i^{-\gamma}$, $1\leq i\leq 500$. 
The parameter $\gamma$ controls the `incoherence' of the matrix, i.e., larger values of $\gamma$ makes the data more `spiky'. Table \ref{fig:lev_comp_power} lists the quality of sparse sketches produced via the two sampling methods. 
%
%
%  ===============================================
\begin{table}[!h]
\begin{center}
    \begin{tabular}{| c | c | c | c |}
   \hline
 &  $\frac{s}{k(m+n)}$ & hybrid-$(\ell_1,\ell_2)$ &  $p_{lev}$    \\ \hline \hline  
%---
\multirow{2}{*}{$\gamma=0.5$} 
&  3 &   42\% &   58\% \\ \cline{2-4}
&  5 &   31\% &   43\% \\ \hline
%---
\multirow{2}{*}{$\gamma=0.8$} 
&  3 &   15\% &   43\% \\ \cline{2-4}
&  5 &   12\% &   40\% \\ \hline
%%---
\multirow{2}{*}{$\gamma=1.0$} 
&  3 &   \text{ }\text{ }8\% &   42\% \\ \cline{2-4}
&  5 &   \text{ }\text{ }6\% &   39\% \\ \hline
%%---
%%$\gamma = 0.8, \frac{s}{k(m+n)}=3$ & \qquad  15\% & \qquad  43\% \\
%%$\gamma = 0.8, \frac{s}{k(m+n)}=5$ & \qquad  12\% & \qquad  40\% \\ \hline
%%$\gamma = 1.0, \frac{s}{k(m+n)}=3$ & \qquad  8\% & \qquad  42\% \\ \hline
%%$\gamma = 1.0, \frac{s}{k(m+n)}=5$ & \qquad  6\% & \qquad  39\% \\ \hline
   \end{tabular}
\caption{
Sparsification quality $\TsNorm{\matA_{pow} - \tilde{\matA}_{pow}}/\TsNorm{\matA_{pow}}$  for low-rank `power-law' matrix $\matA_{pow}$ ($k=5$). We compare the quality of hybrid-$(\ell_1,\ell_2)$ sampling and leverage score sampling for two sample sizes. We note (average) $\alpha^*$ of hybrid-$(\ell_1,\ell_2)$ distribution for data $\matA_{pow}$ using $\epsilon=0.05, \delta=0.1$. For $\gamma=0.5, 0.8, 1.0$, we have $\alpha^* = 0.11, 0.72, 0.8$, respectively. 
}
\label{fig:lev_comp_power}
\end{center}
\end{table}
%  ===============================================

\begin{figure}[!h]
\centering
%\begin{subfigure}
\begin{subfigure}[b]{0.4\textwidth}
	\includegraphics[width=\textwidth,height=3.0cm,width=6.0cm]{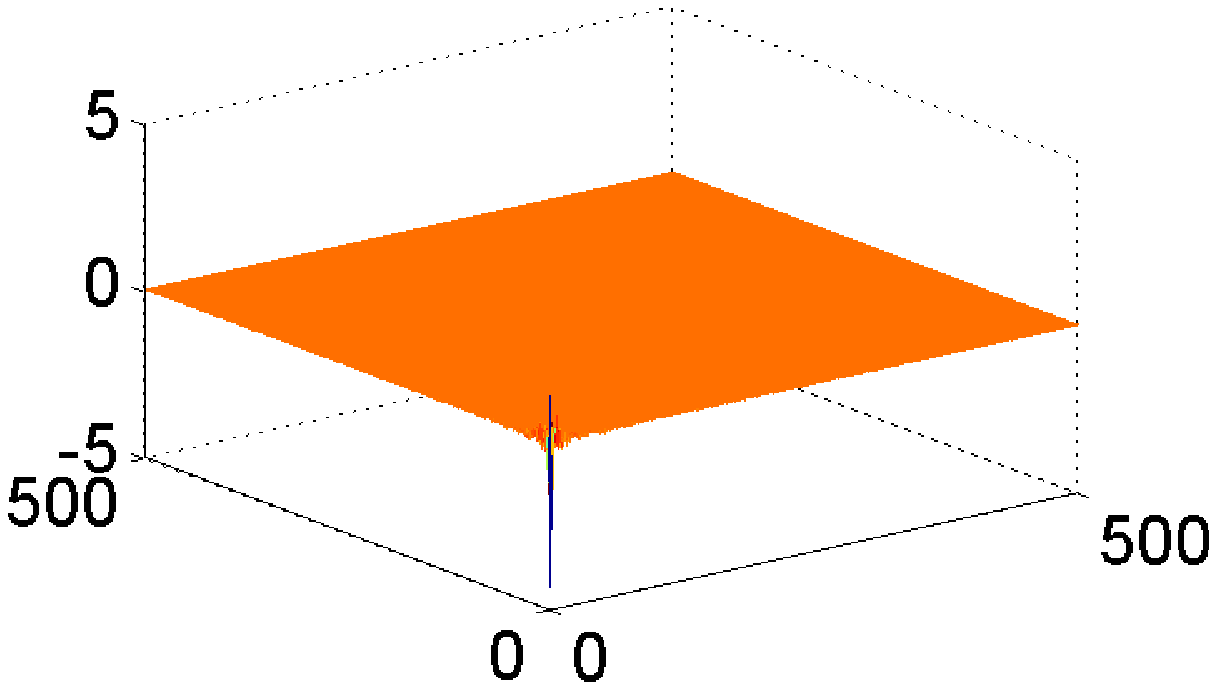}
\caption{Low-rank data $\matA_{pow}$}
\end{subfigure}
\begin{subfigure}[b]{0.4\textwidth}
	\includegraphics[width=\textwidth,height=3.0cm,width=6.0cm]{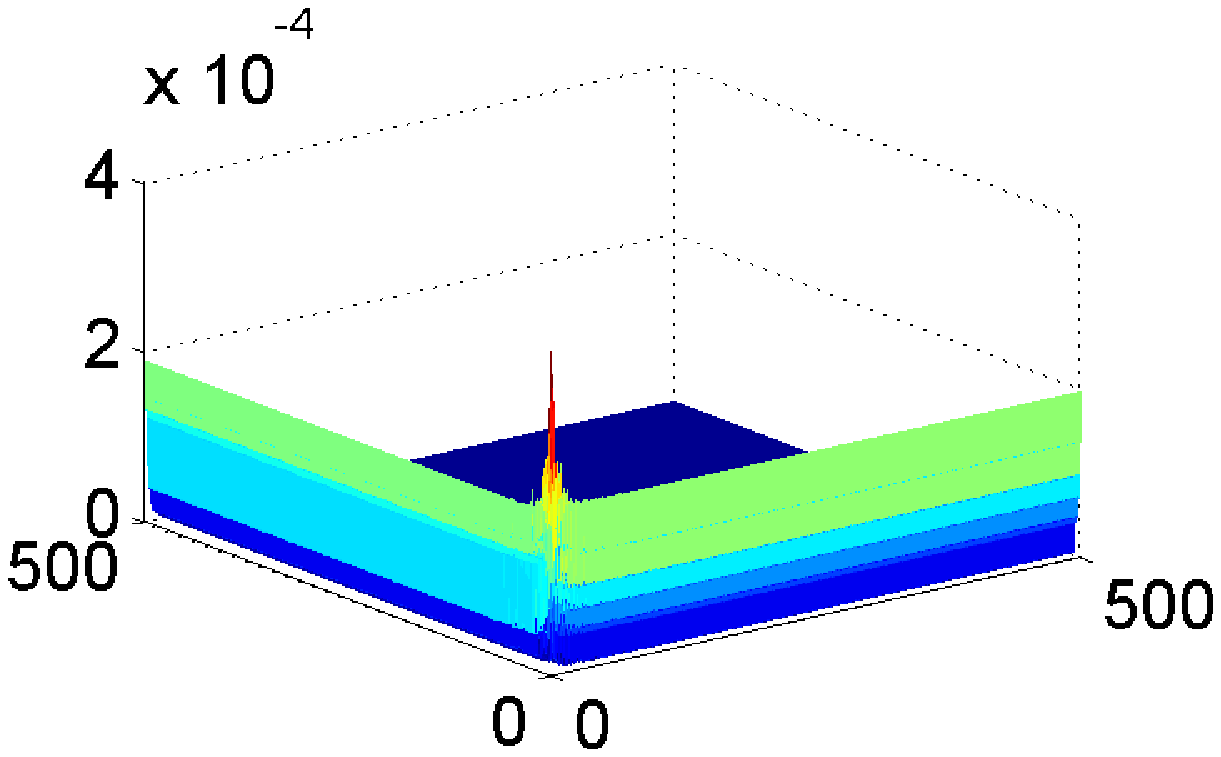}
\caption{Element-wise leverage scores for $\matA_{pow}$}
\end{subfigure}
\begin{subfigure}[b]{0.4\textwidth}
	\includegraphics[width=\textwidth,height=3.0cm,width=6.0cm]{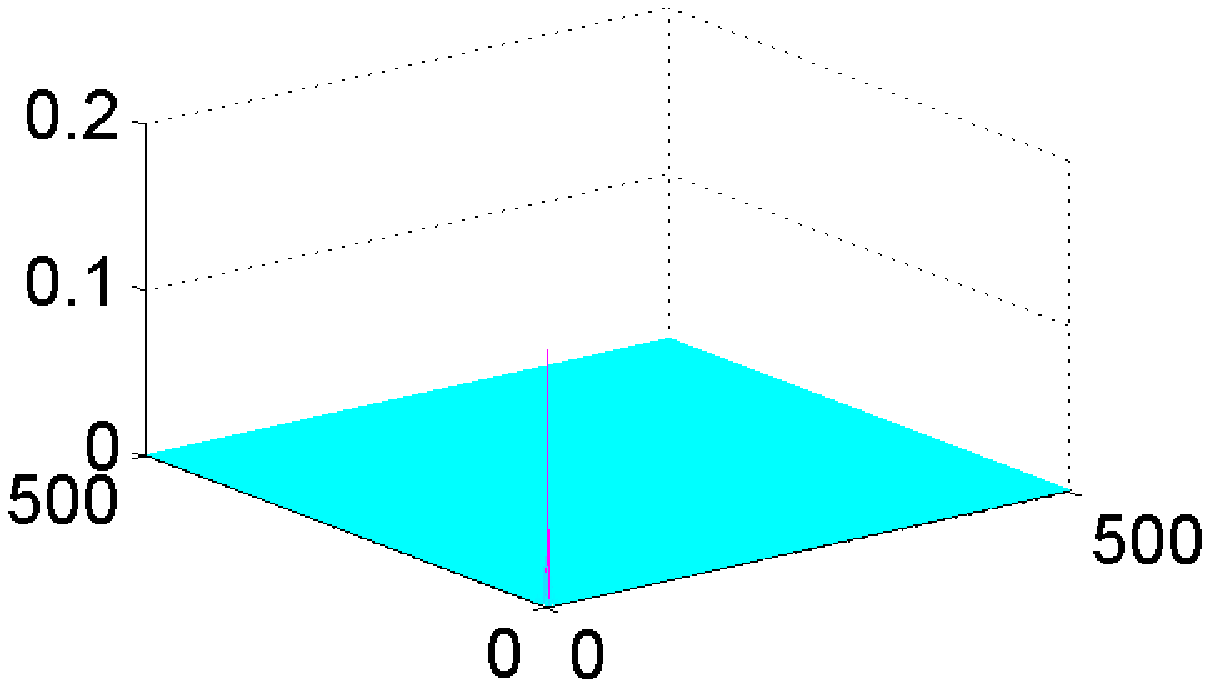}
\caption{Optimal hybrid distribution for $\matA_{pow}$}
\end{subfigure}
\caption{
Comparing optimal hybrid-$(\ell_1,\ell_2)$ distribution with leverage scores $p_{lev}$ for data $\matA_{pow}$ for $\gamma=1.0$. 
(a) Structure of $\matA_{pow}$, (b) distribution $p_{lev}$, (c) optimal hybrid-$(\ell_1,\ell_2)$ distribution. 
Our optimal hybrid distribution is more aligned with the structure of the data, requiring much smaller sample size to achieve a given accuracy of sparsification. This is supported by Table \ref{fig:lev_comp_power}.
}
\label{fig:power_data}
\end{figure}

We note that, with increasing $\gamma$ leverage scores get more aligned with the structure of the data, resulting in gradually improving approximation quality, for the same sample size.
Larger $\gamma$ produces more variance in data elements. $\ell_2$ component of our hybrid distribution bias us towards the larger data elements, while $\ell_1$ works as a regularizer to maintain the variance of the sampled (and rescaled) elements. With increasing $\gamma$ we need more regularization to counter the problem of rescaling.
Interestingly, our optimal parameter $\alpha^*$ adapts itself with this changing structure of data, e.g. for $\gamma=0.5, 0.8, 1.0$, we have $\alpha^* = 0.11, 0.72, 0.8$, respectively.  This shows the benefit of our parameterized hybrid distribution to achieve a superior approximation quality. Figure \ref{fig:power_data} shows the structure of the data $\matA_{pow}$ for $\gamma=1.0$ along with the optimal hybrid-$(\ell_1,\ell_2)$ distribution and leverage score distribution $p_{lev}$. The figure suggests our optimal hybrid distribution is better aligned with the structure of the data, requiring smaller sample size to achieve a desired sparsification accuracy.
%Our optimal hybrid-$(\ell_1,\ell_2)$ turns out to be superior.

We also compare the performance of the two sampling methods, optimal hybrid and leverage scores, on rank-truncated Digit data. It turns out that projection of Digit data onto top three principal components preserve the separation of digit categories. Therefore, we rank-truncate Digit data via SVD using rank three. Table \ref{fig:lev_comp_digit} shows the superior quality of sparse sketches produced via optimal hybrid-$(\ell_1,\ell_2)$ sampling for this rank-truncated digit data.
%  ===============================================
\begin{table}[!h]
\begin{center}
    \begin{tabular}{|  c | c | c |}
   \hline
 & Hybrid-$(\ell_1,\ell_2)$ &  $p_{lev}$    \\ \hline \hline  
$\frac{s}{k(m+n)}=3$ &   44\% &   61\% \\ \hline
$\frac{s}{k(m+n)}=5$ &  34\%  &  47\% \\ \hline
   \end{tabular}
\caption{
Sparsification quality $\TsNorm{\matA - \tilde{\matA}}/\TsNorm{\matA}$ for rank-truncated Digit matrix ($k=3$). We compare the optimal hybrid-$(\ell_1,\ell_2)$ sampling 
%($\alpha^*  = 0.2$ using $\epsilon=0.05, \delta=0.1$) 
and leverage score sampling for two sample sizes.
}
\label{fig:lev_comp_digit}
\end{center}
\end{table}
%  ===============================================

Finally, Table \ref{fig:lev_comp_synth} shows the superiority of optimal hybrid-$(\ell_1,\ell_2)$ sampling for rank-truncated (rank 5)  $\matA_{0.1}$ matrix for matrix sparsification.
%  ===============================================
\begin{table}[!h]
\begin{center}
    \begin{tabular}{|  c | c | c |}
   \hline
 & Hybrid-$(\ell_1,\ell_2)$ &  $p_{lev}$    \\ \hline \hline  
$\frac{s}{k(m+n)}=3$ &   25\% &   80\% \\ \hline
$\frac{s}{k(m+n)}=5$ &  21\%  &  62\% \\ \hline
   \end{tabular}
\caption{
Sparsification quality $\TsNorm{\matA - \tilde{\matA}}/\TsNorm{\matA}$ for rank-truncated $\matA_{0.1}$ matrix ($k=5$). We compare the  optimal hybrid-$(\ell_1,\ell_2)$ sampling 
%(average $\alpha^*  = 0.71$ using $\epsilon=0.05, \delta=0.1$) 
and leverage score sampling using two sample sizes.
}
\label{fig:lev_comp_synth}
\end{center}
\end{table}
%  ===============================================

%
%  ===============================================
% ============================================

% ============================================
% ============================================
\subsubsection{Quality of Fast PCA}

\begin{figure}[!t]
\centering
\begin{subfigure}[b]{0.4\textwidth}
	\includegraphics[height=4.3cm,width=6.5cm]{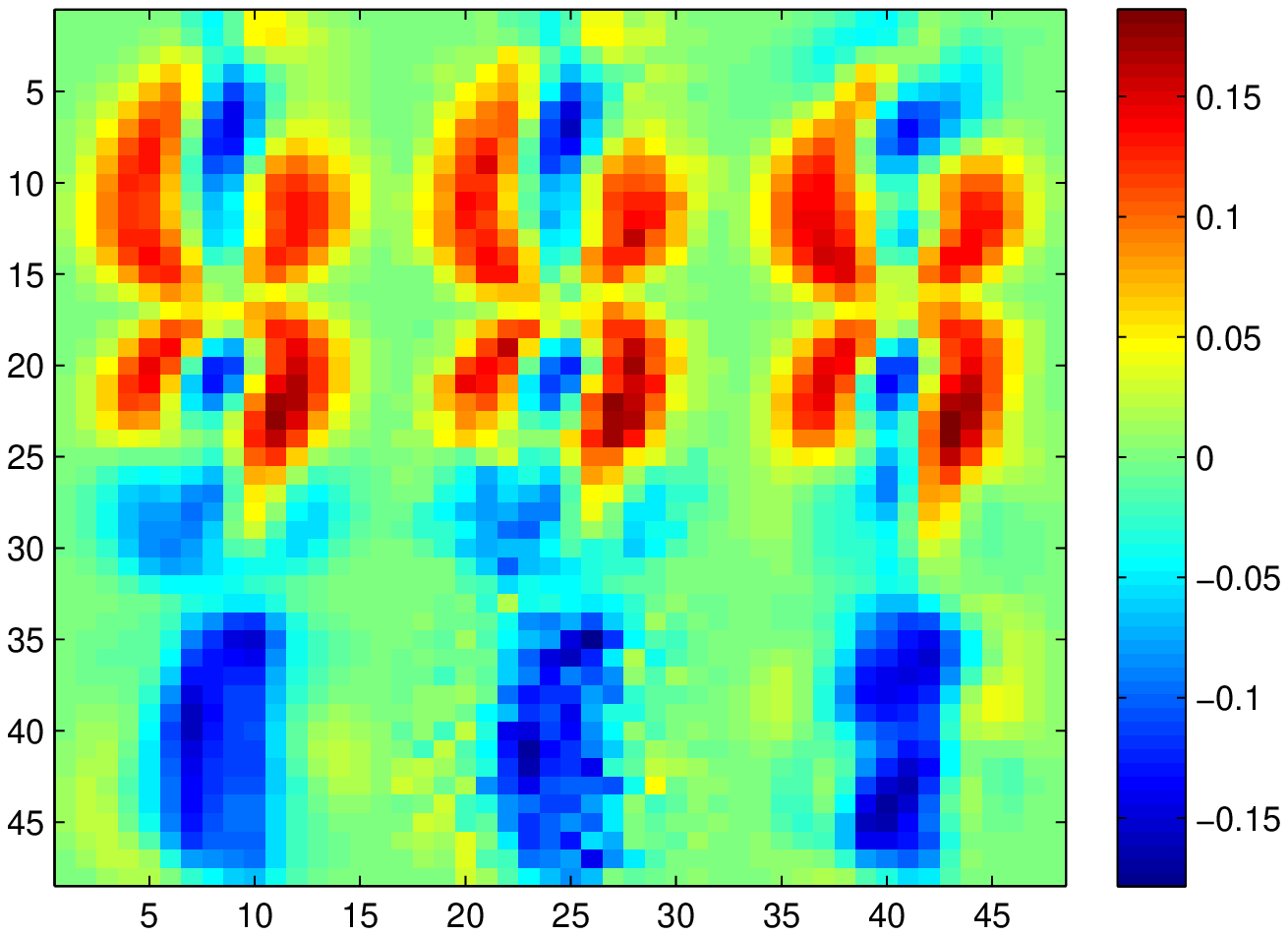}
\caption{PCA's}
\label{fig:digit_PCA}
\end{subfigure}
\quad
\begin{subfigure}[b]{0.4\textwidth}
	\includegraphics[height=4.3cm,width=6.5cm]{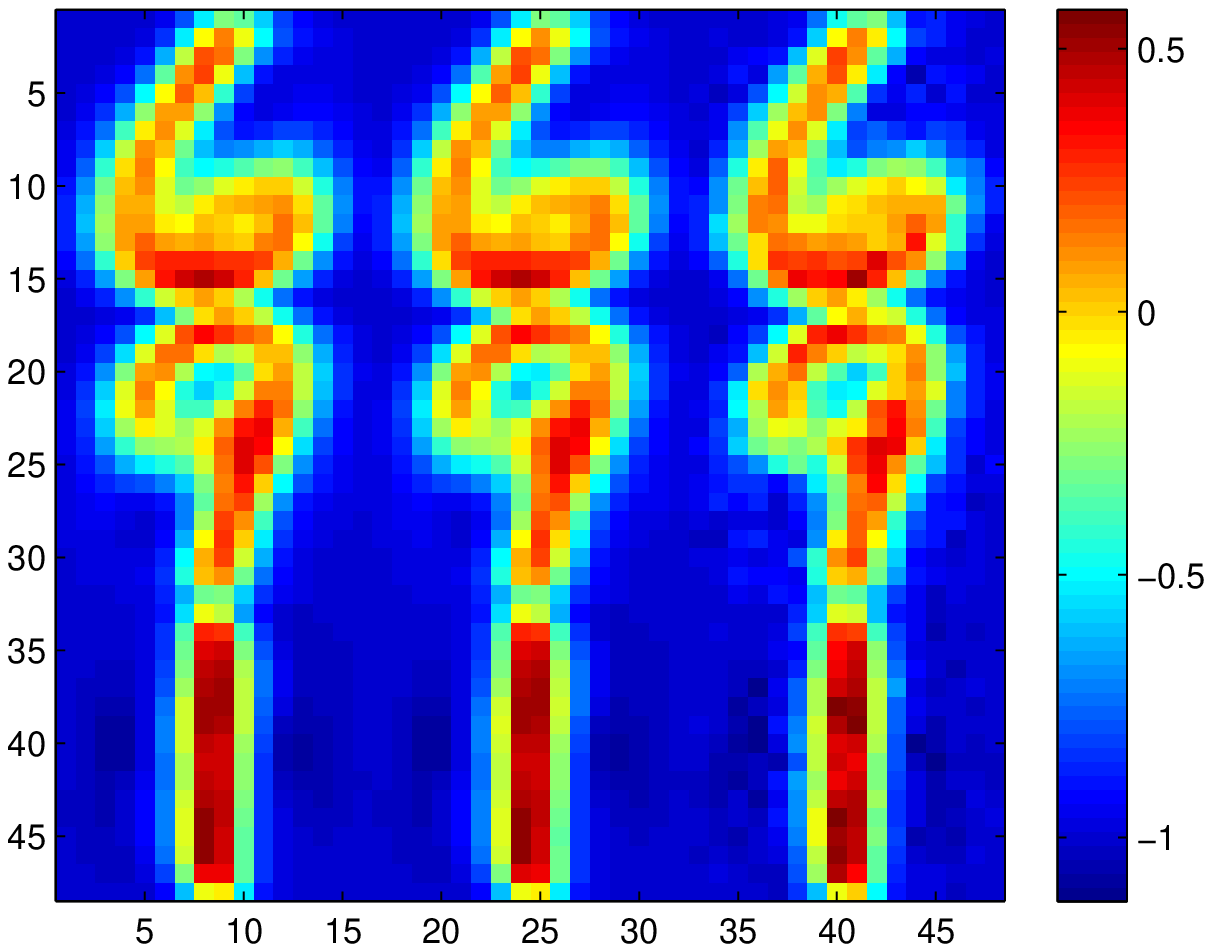}
\caption{Projected data onto the PCA's}
\label{fig:digit_PCA_project}
\end{subfigure}
\caption{
Approximation quality of fast PCA (Algorithm \ref{alg:alg_approx_PCA}) on Digit data.
(a) Visualization of principal components as $16 \times 16$ image. Principal components are ordered from the top row to the bottom. First column of PCA's are exact $\mathcal{A}$. Second column of PCA's are $\mathcal{H}$ computed on sparsified data using $\sim$ 7\% of all the elements via optimal hybrid sampling. Third column of PCA's are $\mathcal{G}$ computed on $\matA_G$. Visually, $\mathcal{H}$ is closer to $\mathcal{A}$.
%, especially for the top two PCA's.
%
(b) Visualization of projected data onto top three PCA's. First column shows the average digits of projected actual data onto the exact PCA's $\mathcal{A}$. Second column is the average digits of projected actual data onto approximate PCA's (of sampled data) $\mathcal{H}$. We observe a similar quality of average digits of projected actual data onto approximate PCA's $\mathcal{G}$ of $\matA_G$. Third column shows the average digits for projected sparsified data onto approximate PCA's  $\mathcal{H}$. 
}
\label{fig:PCA_digit}
\end{figure}

\begin{figure}[!h]
\centering
	\includegraphics[width=0.7\textwidth, height=4.3cm,width=15.2cm]{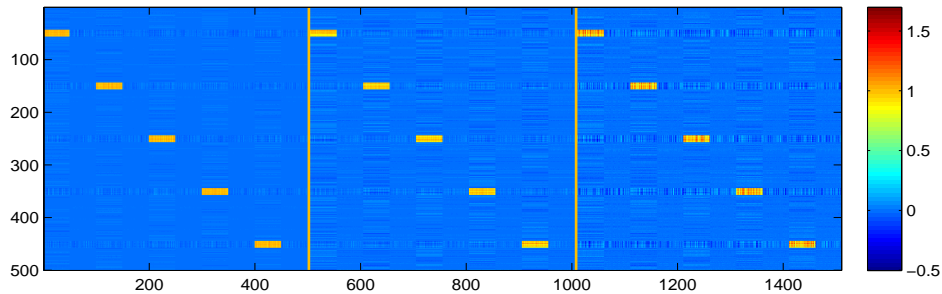}
\caption{
Approximation quality of fast PCA (Algorithm \ref{alg:alg_approx_PCA}) for data $\matA_{0.1}$. Visualization of projected data onto top five PCA's. Left image  shows the projected actual data onto the exact PCAs $\mathcal{A}$. Middle image is the projection of actual data onto approximate PCA's (of sampled data) $\mathcal{H}$. We observe a similar quality of PCA's $\mathcal{G}$ for $\matA_G$. Right image shows the projected sparsified data onto approximate PCA's $\mathcal{H}$. We use only 6\% of all the elements to produce the sparse sketches via optimal hybrid sampling.
}
\label{fig:PCA_synth}
\end{figure}

We investigate the quality of fast PCA approximation (Algorithm \ref{alg:alg_approx_PCA}) for Digit data and $\matA_{0.1}$. 
%With digit data, one can visualize the quality of approximation. For comparison purpose, we create a matrix $\matA_G = \matG\matA$, where $\matA$ is the actual data, and $\matG$ is a $r \times m$ standard Gaussian matrix. 
We set $r=30\cdot k$ for the random projection matrix $\matA_G$ to achieve a comparable runtime of $\mathcal{G}$ with $\mathcal{H}$.
Figure \ref{fig:digit_PCA} shows the PCA (exact and approximate) for Digit data. Also, we consider visualization of the projected data onto top three principal components (exact and approximate) in Figure \ref{fig:digit_PCA_project}. In Figure \ref{fig:digit_PCA_project}, we form an average digit for each digit category by taking the average of pixel intensities in the projected data over all the digit samples in each category. Similarly, Figure \ref{fig:PCA_synth} shows the visual results for data  $\matA_{0.1}$ (we set $k=5$).
Finally, Table \ref{fig:speedup} lists the gain in computation time for Algorithm \ref{alg:alg_approx_PCA} due to sparsification. 
%We define speed up as $(T_{e} - T_{s})/T_{e}$, where 
%Let $T_{e}$, $T_{s}$, and $T_{G}$ be the run time (in milliseconds) of PCA on actual data $\matA$, sparsified data $\tilde{\matA}$, and $\matA_G$, respectively.
%
% ============================================

% ============================================
% ============================================
%  ===============================================
\begin{table}[!h]
\begin{center}
    \begin{tabular}{|  c || c | c |}
   \hline
 & Sparsified Digit &  Sparsified $\matA_{0.1}$   \\ \hline \hline  
Sparsity &   93\% &    94\% \\ \hline
$T_{h}/T_{a}/T_G$ &  30/151/36  &  18/73/36\\ \hline
%Speed up & \qquad 80\% & \qquad \quad 75\% \\ \hline
%
% \qquad \quad & Sparsity& \quad Speed up    \\ \hline \hline  
%Sparsified Digit & \qquad  \quad 93\% & \quad \qquad  94\% \\ \hline
%Sparsified $\matA0.1$ & \qquad \quad 75\% & \qquad \quad 63\% \\ \hline
   \end{tabular}
\caption{
Computational gain of Algorithm \ref{alg:alg_approx_PCA} comparing to exact PCA. We report the computation time of MATLAB function `svds($\matA$,$k$)' for actual data ($T_a$), sparsified data ($T_h$), and random projection data $\matA_G$ ($T_G$). We use only 7\% and 6\% of all the elements of Digit data and $\matA_{0.1}$, respectively, to construct respective sparse sketches.
}
\label{fig:speedup}
\end{center}
\end{table}
% ============= END OF COMMENTED OUT =========================

\subsection{Conclusion}

Overall, the experimental results demonstrate the quality of the algorithms presented here, indicating the superiority of our 
approach to other extreme choices of element-wise sampling methods, such as, $\ell_1$ and $\ell_2$ sampling. Also, we demonstrate the theoretical and practical usefulness of hybrid-$(\ell_1,\ell_2)$ sampling for fundamental data analysis tasks such as fast computation of PCA. Finally, our method outperforms element-wise leverage scores for the sparsification of various \textit{low-rank} synthetic and real data matrices.
%

%-----------------------------------------------------------------------

% \bibliographystyle{plainnat}
%\bibliography{sparsification} \bibliographystyle{plainnat}
%\bibliographystyle{plain}
%-----------------------------------------------------------------------
%\clearpage \input{appendix}
\end{document}